----------
X-Sun-Data-Type: default
X-Sun-Data-Description: default
X-Sun-Data-Name: ind12n.tex
X-Sun-Content-Lines: 2055

\documentstyle[12pt]{article}
 \input amssym.def
\input amssym

\title{Induced modules for vertex operator algebras
\footnotetext{The first author was supported by NSF grant
DMS-9303374 and a research grant from the Committee on Research, UC
Santa Cruz.} \footnotetext{The second author was supported by NSF
grant DMS--9401389.} }
\author{Chongying Dong and  Zongzhu Lin}
\date{}

\makeatletter
\def\@begintheorem#1#2{\sl \trivlist \item[\hskip \labelsep{\bf #1 }]}
\def\@opargbegintheorem#1#2#3{\sl \trivlist \item[\hskip \labelsep{\bf #1\
(#3)}]}

\newtheorem{th}{Theorem}
\newtheorem{prop}{Proposition}
\newtheorem{lem}{Lemma}
\newtheorem{cor}{Corollary}

\newcommand{\text}[1]{\mbox{#1}}\newcommand{\rom}[1]{{\rm #1}}
\newcommand{\qed}{\mbox{ $\square$}}\newcommand{\pf}{\noindent {\em Proof: \,}}

\newcommand{\comb}[2]{{\renewcommand\arraystretch{0.7}
\!\left(\!\begin{array}{c}\! {\scriptstyle #1}\!\\ \!{\scriptstyle
#2}\!\end{array}\!\right)\!}}

\def \Z{\Bbb Z}
\def \C{\Bbb C}
\def \R{\Bbb R}
\def \Q{\Bbb Q}
\def \N{\Bbb N}
\def \<{\langle}
\def \>{\rangle}
\def \wt{\text{\rom{wt}}}

\def \Res{\text{\rom{Res}}}

\def \End{\text{\rom{End}}}

\def \Ind {\text{\rom{Ind}}}

\def \M{\text{\rom{Mod}}}

\def \LL{{\cal L}}
\def \CC{{\cal C}}
\def \I{{\cal I}}
\def \X{{\cal H}}
\def \F{{\cal F}}

\def \Ker{\text{\rom{Ker}}}

\def \an{\text{\rom{Ann}}}
\def \ext{\text{\rom{Ext}}}
\def \Hom{\text{\rom{Hom}}}
\def \ind{\text{\rom{Ind}}}
\def \l{\lambda}

\newcommand{\rem}{\smallskip \noindent {\em Remark. \, }}

\newcommand{\blist}{\begin{list}{\text{\rom{(\roman{enumii})}}}{\setlength{\leftmargin}{2em}
\setlength{\itemindent}{0ex}
\setlength{\labelsep}{2ex}\setlength{\listparindent}{\parindent}
\usecounter{enumii}}}
\newcommand{\elist}{\end{list}}
\newcommand{\bsub}{\begin{list}{\bf
\arabic{section}.\arabic{subsection}.}{\setlength{\leftmargin}{0em}
\setlength{\itemindent}{7ex}
\setlength{\labelsep}{2ex}\setlength{\listparindent}{\parindent}
\setlength{\itemsep}{10pt}
\usecounter{subsection}} }
\newcommand{\esub}{\end{list}}
\newcommand{\subsect}[1]{\item{\bf #1}}

\begin{document}
\maketitle
\setcounter{secnumdepth}{4}


\begin{abstract}
 For a vertex operator algebra $V$ and a vertex operator
subalgebra
$V'$ which is invariant under an automorphism $g$ of $V$
of finite order, we introduce a $g$-twisted induction functor from
the category of $g$-twisted $V'$-modules to the category
of $g$-twisted $V$-modules. This functor satisfies
the Frobenius reciprocity and transitivity. The results are illustrated
with $V'$ being the $g$-invariants in simple $V$ or $ V'$ being $g$-rational.
\end{abstract}
\section{Introduction}

A lot of progress on the representation theory for vertex operator algebras
has been made in the last few years.
For example, the representation theory for the concrete vertex operator
algebras, which include
the moonshine vertex operator algebra $V^{\natural}$ ([FLM],[D3]),
the vertex operator algebras based on even positive definite lattices [D1],
the vertex operator algebras associated with the integrable representations
of affine Lie algebras and Virasoro algebras ([DMZ], [DL], [FZ], [W]), have
been studied extensively. There are also abstract approaches such as
Zhu's one to one correspondence between the set of
inequivalent irreducible modules for a
given vertex operator algebra and  the set of inequivalent
irreducible modules for an
associative algebra associated with the vertex operator algebra [Z],
and the tensor products of modules ([HL] and [L]); See also [FHL]
for the results concerning intertwining operators and contragredient
modules.
Many of these results are analogues of the corresponding results in
the classical Lie algebra theory.

The purpose of this paper is to give a construction of induced twisted
modules for vertex operator algebras and present some initial results. The
main idea in constructing the induced module comes from the induction
theory for the representations of Lie groups, algebraic groups,
quantum groups, Hopf algebras (\cite{vogan}, \cite{ja}, \cite{apw},
[Lin1-Lin2]). In order for the induced module to have the functorial
property, one has to
enlarge the category of $g$-twisted modules to ensure the existence
of the induced modules. This resembles Harish-Chandra's
theory in the representation theory of Lie groups.
We prove that in most interesting cases, the induced modules from a simple
module for a vertex operator subalgebra are indeed $g$-twisted modules.
The structures of these modules in certain special cases are discussed.

One of the main motivations for introducing the induced modules is to
study the ``orbifold conformal field theory.''
Roughly speaking, an orbifold theory
is a conformal field theory which is obtained from a given conformal field
theory modulo the action of a finite symmetry group (see [DVVV]).
Let $V$ be a vertex operator algebra and $G$ be a finite
subgroup of the automorphisms of $V.$ Denote by $V^G$ the subspace of
$V$ consisting of the fixed points under the action of $G.$
Then $V^G$ is a vertex operator subalgebra of $V.$
Algebraically, the orbifold theory is to study the representation theory
of $V^G.$  The main new feature of the orbifold theory is the introduction of
twisted modules. A $g$-twisted module for $g\in G$ is automatically
an ordinary module for $V^G$ under the restriction. It is proved in
[DM2] that if $V$ is holomorphic and $G$ is nilpotent then
any irreducible $g$-twisted $V$-module is completely reducible as a
$V^G$-module. It is conjectured that this is true for arbitrary $G$
and that any irreducible $V^G$-module appears as an irreducible component
of some irreducible twisted module (see e.g. [DVVV], [DPR] and [DM2]).
The theory of induced twisted modules  for vertex operator
algebras, discussed in this paper, is being developed with this conjecture
in mind.

This paper is organized as follows: In Section 2, after
recalling the notion of twisted module for a vertex operator algebra from
[D2] and [FFR], we define the twisted enveloping algebra $A(g).$
We give a necessary and sufficient condition under which an $A(g)$-module
is a $g$-twisted module.  A linear topology is defined on $A(g)$ by
a topological basis at 0 consisting of all $\an_{A(g)}(m)$ for
  an element $m$ in a twisted module. It turns out this topology is the weakest
topology
on $A(g)$ so that the $A(g)$-module structure on any twisted module
$M$ gives a continuous map from $A(g)$ to $\End_{\C}(M)$, which is equipped
with the point-wise convergence topology. It is important to note that we  need
the
representations to be continuous in order for the Jacobi identity
to hold.  We also introduce a certain $A(g)$-module category $\bar {\cal
C}_g$ which contains the $g$-twisted
$V$-module category ${\cal C}_g$ as a full subcategory. In fact,
$\bar{\cal C}_g$ consists of the objects of ${\cal C}_g$ together with
their direct limits in the category of $ A(g)$-modules.

Section 3 which devotes the definition of induced module
is the center of the paper. Let $g$ be a finite order automorphism of $V$
and $V'$ a subalgebra of $V$ which  is $g$-invariant. Denote the restriction
of $g$ to $V'$ by  $g'.$  Then   there is an
algebra embedding from $A(g')$  into $A(g).$  For a $g'$-twisted $V'$-module
$W$ we define $\Ind_{\bar {\cal C}'_{g'}}^{\bar{\cal C}_g}(W)$
to be the subspace  of
$\Hom_{A(g')}(A(g),W)$
consisting of elements which are killed by some
$\an_{A(g)}(m)$ for some $m$ in a $g$-twisted
$V$-module. Then we prove that this induction functor enjoys all the
properties  of  an induction functor  should have, such as
the Frobenius reciprocity and the transitivity.

In Section 4 we investigate the $g$-induced modules for vertex operator
algebra with only finitely many irreducible $g$-twisted modules.
 In the case that
$V$ is $g$-rational, that is, any $g$-twisted module is completely reducible,
we show that the induced module of $W$, which has a composition series,
{}from a $g$-invariant subalgebra to $V$ is in fact in ${\cal C}_g$.
 In particular, the induced module from  any irreducible
module in this case is an ordinary $g$-twisted module.

In Section  5 we    discuss the  induction from a subalgebra which is
the fixed points  of an finite automorphism. First we show that
the cyclic group $\<g\>$ acts on any irreducible $g$-twisted module,
each eigenspace of $g$ is an irreducible $V^G$-module, different eigenspaces
are inequivalent $V^G$-modules, and the eigenspaces from inequivalent
irreducible $g$-twisted modules are inequivalent $V^G$-modules. Using
this result together  with Frobenius reciprocity  we show that
the $g$-induced module of an  eigenspace of $g$ in
an irreducible $g$-twisted
module $W$ from $V^G$ to $V$ has a unique simple  $g$-twisted submodule
isomorphic to $W,$ and the $g$-induced modules from any  other irreducible
$V^G $ module is zero.

\section{Twisted enveloping algebras and their modules.}

\label{sec1}
\bsub

\subsect{Twisted modules.}\label{sec1.1}
Let $(V,Y,{\bf 1},\omega)$ be a vertex operator
algebra (we refer to [B], \cite{fhl} and \cite{flm} for definitions and
properties)
 and  $g$ be an automorphism of $V$ of finite order $K.$  Then $V$ is a direct
sum of eigenspaces of $g:$
$$V=\oplus_{r\in \Z/K\Z}V^r$$
where $V^r=\{v\in V|gv=e^{r2\pi i/K}v\}$.
(We will use $ r \in \{0, 1,\ldots, K-1\}$ to denote both
the representing residue class and the integer itself.)
Following [D2] and [FFR], a $g$-twisted
$module$ $M$ for $V$ is a $\C$-graded
vector space:
$$M=\coprod_{\lambda \in{\C}}M_{\lambda} $$
(for $ w\in M_{\lambda}$, $\lambda =\mbox{wt}\,w $ is
called the weight of $w$) such that for each $\lambda \in \C $
\begin{equation}
\dim M_{\lambda}<\infty\; \mbox{ and } M_{n+\lambda}=0\; \mbox{ for }
n \in {\Bbb Z} \ \mbox{sufficiently small};
\label{bound1}
\end{equation}
and  equipped with a linear map
$$\begin{array}{l}
V\to (\mbox{End}\,M)\{z\}\label{map}\\
v\mapsto\displaystyle{ Y_M(v,z)=\sum_{n\in\Q}v_nz^{-n-1}\ \ \ (v_n\in
\mbox{End}\,M)}
\end{array}$$
(where for any vector space $W,$ we define $W\{z\}$ to be the vector
space of $W$-valued formal series in $z,$ with arbitrary real powers of $z$)
satisfying the following conditions for $u,v\in V$,
$w\in M$, and $r\in\Z/K\Z$:
\begin{eqnarray}
& &Y_M(v,z)=\sum_{n\in \frac{r}{K}+\Z}v_nz^{-n-1}\ \ \ \ \text{\rom{for}}\ \
v\in V^r;\label{1/2}\\
& &v_lw=0\ \ \  					
\mbox{for}\ \ \ l\in \Q \ \ \mbox{sufficiently\ large};\label{vlw0}\\
& &Y_M({\bf 1},z)=1;\label{vacuum}
\end{eqnarray}
 \begin{equation}\label{jacobi}
\begin{array}{c}
\displaystyle{z^{-1}_0\delta\left(\frac{z_1-z_2}{z_0}\right)
Y_M(u,z_1)Y_M(v,z_2)-z^{-1}_0\delta\left(\frac{z_2-z_1}{-z_0}\right)
Y_M(v,z_2)Y_M(u,z_1)}\\
\displaystyle{=z_2^{-1}\left(\frac{z_1-z_0}{z_2}\right)^{-r/K}
\delta\left(\frac{z_1-z_0}{z_2}\right)
Y_M(Y(u,z_0)v,z_2)}
\end{array}
\end{equation}
if $u\in V^r;$
$$[L(m),L(n)]=(m-n)L(m+n)+\frac{1}{12}(m^3-m)\delta_{m+n,0}(\mbox{rank}\,V)$$
for $m, n\in {\Z},$ where
\begin{eqnarray}
& &L(n)=\omega_{n+1}\ \ \ \mbox{for}\ \ \ n\in{\Z}, \ \ \
\mbox{i.e.},\ \ \ Y_M(\omega,z)=\sum_{n\in{\Z}}L(n)z^{-n-2};\nonumber\\
& &L(0)w=nw=(\mbox{wt}\,w)w \ \ \ \mbox{for}\ \ \ w\in M_n\
(n\in{\C});\label{6.71}\\
& &\frac{d}{dz}Y_M(v,z)=Y_M(L(-1)v,z).\label{6.72}
\end{eqnarray}
This completes the definition. We denote this module by
$(M,Y_M)$ (or briefly by $M$).

The following are consequences of the definition
\begin{eqnarray}
& &[L(-1),Y_M(v,z)]=Y_M(L(-1)v,z);\label{6.73}\\
& &[L(0),Y_M(v,z)]=Y_M(L(0)v,z)+zY_M(L(-1)v,z).\label{l0}
\end{eqnarray}

\rem From (\ref{jacobi}), (\ref{6.71}),
(\ref{6.72}) and (\ref{l0}) we find
that if $v\in V$ is homogeneous, then $v_n$ has weight wt\,$v-n-1$ as
an operator, that is, $v_n$: $M_{\lambda}\to M_{\lambda+\rm{wt}\,v-n-1}.$
In particular, $L(n)$ has weight $-n.$

\rem  For each $ v \in V $, and $ u \in V^{r}$ and
any $ p \in \Z $ and $ s, t  \in \R $, we can compare the coefficients of
$ z_0^{-p-1}z_{1}^{-s-1}z_{2}^{-t-1}$ on the both sides of the Jacobi
identity (\ref{jacobi}) to get
\begin{equation}
\sum_{m\geq 0}(-1)^m \comb{p}{m}(u_{p+s-m}v_{t+m}-(-1)^{p}v_{p+t-m}u_{s+m})=
\sum_{m\geq 0}\comb{s}{m}(u_{p+m}v)_{s+t-m}.
\label{jacobi-term}
\end{equation}
On the other hand,
all these identities together will imply the Jacobi identity
(\ref{jacobi}).  We will use these identities more often.

A vertex operator algebra $V$ is called $g$-quasi-rational if $V$ has only
finitely many irreducible $g$-twisted modules. If $V$ further satisfies the
condition that any $g$-twisted module is completely reducible, we call
$V$ a $g$-rational vertex operator algebra. If $g=1,$ we get a rational vertex
operator algebra.

\subsect{}\label{sec1.2}
Let $M $ and $N$ be two $ g $-twisted $ V $-modules.
A homomorphism $  \phi$: $M \to N $ of $ g $-twisted $V $-modules is a
linear map which commutes with the operators $
v_{n} $ for all $ v \in V $ and $ n \in \frac{1}{K}\Z  $.
In particular, a homomorphism is always homogeneous homomorphism of
graded
vector spaces in the sense that $ \phi(M_{\lambda})\subseteq N_{\lambda}$
since $ \phi $ commutes with the  operator $ L(0)$.

Let ${\cal C}_g(V)$ be the category
of $g$-twisted $V$-modules. Then $ {\cal C}_{g}(V) $ is an abelian
category. For simplicity, we will use $ {\cal C}_{g} $  if the vertex operator
algebra $ V $ is understood. We will abuse the notation a
little by using $M \in {\cal C}_{g} $ to stand for $ M $ being  an object
of the category $ {\cal C}_{g} $.

Let $ V' $ be another  vertex operator algebra and
$ \phi$: $V' \to V $ a homomorphism of vertex operator algebras.
If $ g' $ is an automorphism of $ V' $ such that $ \phi \circ g' =
g\circ \phi $, then $ \phi $ sends the $ g' $-eigenspaces to $ g $-eigenspaces.
In fact, let $K'$ be the order of $ g'$. If $v' \in (V')^{r'}$, then
$\phi(v') \in V^{r}$. Here  we have $\frac{r}{K}=\frac{r'}{K'}$.
By examining Jacobi identity, one can verify that
 every $g$-twisted $V$-module
is automatically a $g'$-twisted $V'$-module.
In this way, $ \phi $ induces a functor $ {\cal C}_{g} \to
{\cal C}_{g'}$, which is exact.

In the sequel, we will describe the category
$ {\cal C}_{g} $ as a  full subcategory of the category of all modules
for an associated algebra.

\subsect{$g$-twisted enveloping algebras.}\label{sec1.3}
Consider the $\C$-vector space
$\bar V=\oplus_{i=0}^{K-1}(V^i\otimes t^{i/K}\C[t,t^{-1}])$ and the tensor
algebra $T(\bar V).$
Note that $ \bar V $ is a graded vector space with the
gradation $ \bar V = \oplus_{ \mu\in \frac{1}{K}\Z} (\bar V)_{\mu}$.
Here $ (\bar V)_{\mu}=\oplus_{n\in \Z}(V_{n}\otimes t^{n-\mu-1})$.
Thus the tensor algebra $ T(\bar V)$
 is  a graded algebra with the gradation
induced by the  gradation on the vector space $ \bar V $.

For each $ g$-twisted $ V$-module $M$, the
module structure of $V$ on $M$ induces a linear map
$$\begin{array}{ccc}
 \bar V & \to & \End_{\C}(M)\\
 v\otimes t^n& \mapsto &v_n,
\end{array}$$
where $v_n$ is the component operator of  $Y_M(v,z)\!=\!\sum_{n\in
\frac{1}{K}\Z}v_nz^{-n-1},$
which extends uniquely to a homomorphism of associative algebras
$$\rho_M:\ T(\bar V)\to \End(M).$$

Let $ \End_{\C}(M)_{\mu}=\{ f \in \End_{\C}(M)\ \; | \;
f(M_{\lambda})\subseteq M_{\lambda +\mu} \text{\rom{ for all }}
\lambda \in \C  \} $. Then the $ \C $-graded algebra $
\End_{\text{\rom{gr}}}(M)
=\oplus_{ \mu \in \C} \End_{\C}(M)_{\mu}$ is a  subalgebra
of $ \End_{\C}(M)$. By the definition of the homomorphism $ \rho_M$,
one can see that $ \rho_M(T(\bar V)) \subseteq \End_{\text{\rom{gr}}}
(M) $ and the map $ \rho_M$: $T(\bar V) \rightarrow \End_{\text{\rom{gr}}}
(M)$ is a homomorphism of graded algebras. Thus the kernel
$ \Ker(\rho_M) $ is a graded ideal of $ T(\bar V)$.

Set $I=\cap_{M\in{\cal C}_g}\ker\rho_M.$ Then $I$ is also a graded
 ideal of $T(\bar V).$
We define $U_g(V)=T(\bar V)/I $, which is also a
$\frac{1}{K}\Z$-graded algebra with
$$U_g(V)=\oplus_{n\in\frac{1}{K}\Z}(U_g(V))_n$$
 such  that
$$\wt(v\otimes t^n)=\wt\, v_n =\wt\,v-n-1$$
for homogeneous $v\in  V.$ We will also use  $ v_{n} $ to denote the
image of $ v\otimes t^{n} $ in $ U_g(V) $ acting
on any $g$-twisted $ V $-modules.
Consider the Lie algebra $sl_2 $ generated by the operators
$\{ L(1), L(0), L(-1)\}$. Then $ U_g(V) $ becomes an $ sl_2 $-module
with $L(1), L(0) $, and $L(-1)$ acting on $U_g(V)$ as derivations. As $ sl_2
$-module, $U_g(V)$ is a direct sum
of $ L(0)$-weight spaces, which gives $ U_g(V) $ the graded algebra structure.
Furthermore, any $g$-twisted modules
are automatically modules for $ sl_2$ with weight space decomposition
defining the graded vector space structure.

\begin{lem}\label{lem1.2}
(1) Every $ g$-twisted $V$-module $ M $ is naturally
a graded left $ U_{g}(V)$-module with the graded structure given by the
weight spaces decomposition as $ sl_2 $-module.

(2) Any  $ U_{g}(V)$-submodules and
 quotient $ U_{g}(V)$-modules  of a $ g $-twisted $ V $-module  $ M$
are also  $g$-twisted $V$-modules.

(3) For any two $ g $-twisted $ V $-modules $ M, N$, a linear map
$ \phi$: $N \rightarrow M $ is a homomorphism of $ g $-twisted
module if and only if $ \phi $ is a homomorphism of $ U_g(V)$-modules.
\end{lem}

\pf (1) It follows from the
construction of $ U_{g}(V)$. To prove (2), let
$ N \subseteq M $ be a  $ U_{g}(V)$-submodule of $ M $.
We first show that $ N$ is a graded $ U_g(V)$-submodule of
$M$. Let $ x \in N $ and $ x= \sum x_{\lambda}$ be a finite
sum in $ M $ with $ x_{\lambda} \in M_{\lambda}$ for distinct
$\lambda $'s. We must show that $ x_{\lambda} \in N $.
We use induction on the number of the nonzero terms of the summation.
By applying  the operator $ L(0) $ (the image of $ \omega \otimes t^{1}$
in $ U_g(V)$) we get $(L(0)-\lambda_{0})x=
\sum (\lambda-\lambda_{0})x_{\lambda}$ with a shorter expression. By induction
hypothesis, we have
$ x_{\lambda} \in N $. Now
the conditions on the gradation of a $g$-twisted $ V$-module
is automatically satisfied.
If $Y_M(v,z)\!=\!\sum_{n\in \frac{1}{K}\Z}v_nz^{-n-1},$ with $v_n \in
\End(M)$, we have $  v_{n}N \subseteq N $. So
we can consider $v_{n}|_{N} \in \End(N)$ and define
$ Y_{N}(v, z)=\!\sum_{n\in \frac{1}{K}\Z}v_n|_{N}z^{-n-1}$.
Now all other conditions for a $g$-twisted $V$-module
structure on $ N$ can be verified straightforward.
Note that the both sides of the $g$-twisted Jacobi identities
make sense as operators on $N$ at every element of $N$.
For the quotient, since the kernel is a $U_g(V)$-submodule, it is a
$ g $-twisted $ V $-submodule of $ M $. Thus the quotient is a
graded module and one can verify that  the $ U_g(V) $-structure
on the quotient $ g $-twisted $ V $-module $ M/N $ is the
same as  the quotient of the $ U_g(V) $-modules.

(3) It is straightforward to verify that any homomorphism of
$g $-twisted $ V $-modules is naturally a homomorphism of
$ U_g(V) $-modules. Conversely, if a linear map $ \phi $ is a
$ U_g(V)$-homomorphism, then  it has to commute with the operator
$ L(0)$. This shows that $ \phi(N_{\lambda})\subseteq M_{\lambda}$.
Since $ \phi $ commutes with all $ v \otimes t^{n}$, then
we have $ \phi\circ Y_{N}(v, z)= Y_{M}(v, z)\circ\phi $.

\subsect{The Algebra $ A(g)$.} \label{sec:ag}
Note that many of the identities such as the Jacobi
identities  will involve
infinite sum, which does not make any sense in the purely algebraic
setting in $ U_g(V) $.

For each $ \C $-vector space $ A $, let
$ A[[X]] $ be the vector space of formal power series. The natural
embedding  $ A \to  A[[X]] $ (sending elements of $A$ to power
series with constant term only) makes $ A $ a subspace of $ A[[X]]$.
If $A $ is further an algebra, then $ A[[X]] $ is an
algebra containing $ A $  as a subalgebra.
Here the multiplication is given  by the  multiplication of
formal power series.

Let $ M $ be a vector space, and $ A=\End(M) $.
Then $ A[[X]] $ has a subspace
\[C_M=\{f(X)=\sum_{i=0}^{\infty}a_{i}X^{i} \in A[[X]] \; | \;
\sum_{i=0}^{\infty}a_{i}mX^i \in M[X] \text{ for each }m \in M \}.
\]
In fact $ C_M $ is a subalgebra of $A[[X]]$. In this way, we can make
$ M $  a
$ C_M $-module  by defining the action $ f(X)m=\sum_{n=0}^{\infty}a_{n}m $.

Consider the vector space $ U_g(V)_{n}[[X]] $ and the algebra
$ \oplus_{n \in \frac{1}{K}\Z }(U_{g}(V)_{n}[[X]])$, which is a
 proper subalgebra of $ U_g(V)[[X]]$.
For each $ g $-twisted $ V $-module $M $,
the homomorphism $ \rho_M$: $U_g(V) \to \End(M) $ induces a
homomorphism $ \rho_M[X]$:  $\oplus_{n \in
\frac{1}{K}\Z }U_{g}(V)_{n}[[X]] \to \End(M)[[X]] $.

Let $ C= \cap_{M\in {\cal C}_{g}}(\rho_M[X])^{-1}(C_{\End(M)})$.
Then $ C $ is a subalgebra of $  \oplus_{n \in
\frac{1}{K}\Z }U_{g}(V)_{n}[[X]] $.
Note that $ \oplus_{n \in \frac{1}{K}\Z }U_{g}(V)_{n}[[X]] $ is a
$\frac{1}{K}\Z $-graded algebra by defining $\wt X=0$ and $\wt fh=\wt f+\wt h$
for $f,h\in  \oplus_{n \in \frac{1}{K}\Z }U_{g}(V)_{n}[[X]].$
Then $ C=\oplus_{n \in\frac{1}{K}\Z}C_{n} $ is a graded subalgebra
with $ C_{n}=U_{g}(V)_{n}[[X]]\cap C $.

Since $ M $ in $ {\cal C}_{g} $ is a natural graded
$ C $-module, $\an_C(M)$ is an graded (two-sided) ideal of $C.$
Define
$ A(g)=C/(\cap_{M \in {\cal C}_{g}}\an_{C}(M)). $
Then $ A(g) $ is a graded algebra. The constant subalgebra
$ U_{g}(V) $ of $  \oplus_{n \in
\frac{1}{K}\Z }U_{g}(V)_{n}[[X]] $ is actually contained in $ C $.
Moreover we have a natural algebra embedding  $ U_{g}(V) \to A(g)$.
Note that the $sl_2$-module structure on $U_g(V) $ extends naturally
to $ C_M$, which has a weight spaces decomposition and the map $\rho_M$:
$C \to \End(M) $ is a homomorphism of $ sl_2$-modules. Thus the kernels
$\an_{C}(M) $ are graded  ideals. This naturally  induces a graded algebra
structure on $A(g)$.

\subsect{}\label{sec.agm}
By our construction of the algebra $ A(g)$, every $ g $-twisted
$ V $-module extends to an $ A(g) $-module.
We can now regard $ U_g(V) $ as a subalgebra of $ A(g) $.
Thus, every $ g $-twisted $ V $-module $ M $ is a $ A(g)$-module
and every homomorphism between any two $ g $-twisted $ V $-modules
are also a homomorphism of $ A(g)$-modules.
This yields a functor $ {\cal E}$: ${\cal C}_g \to A(g)$-\M\
 (the category of all $ A(g)$-modules).

 \begin{th}
 The functor $ {\cal E} $ is  an isomorphism of $ {\cal C}_{g} $ to a
full subcategory of $ A(g)$-\M.
\end{th}
\pf First we show that $ {\cal E} $ is full.
Let $ M $ and $ N $ be two modules in $ {\cal C}_g $.
Then a linear map $ f$: $M \to N $ is a homomorphism of
$ g $-twisted $ V $-modules if and only if it $ U_g(V) $-linear
and further, if and only if it is $ A(g) $-linear.
This shows that
$ \Hom_{{\cal C}_{g}}(M,N)=\Hom_{A(g)\mbox{-\M}}(M,N) $ as vector    spaces.
This shows that $ {\cal E} $ is full.
The faithfulness and the exactness  follow from the fact that
$ {\cal E} $ does not change the vector space.

Let $ {\cal C}_{A(g)} $ be the full subcategory of $ A(g)$-modules
 satisfying the conditions of (i).
Then (i) defines the functor $ {\cal E}^{-1} $ and this shows that
$ {\cal E} $ is an isomorphism between $ {\cal C}_{g} $
and $ {\cal C}_{A(g)}$.
This proves the theorem.
\qed

 {}From now on, we can simply identify $ {\cal C}_{g} $
with $ {\cal C}_{A(g)}$ and still use $ {\cal C}_{g} $ to denote it.

\subsect{Topology on $A(g).$}\label{sec1.4}
For each vector space $M,$ $\End_{\C}(M)$ is equipped with a  natural product
topology (point-wise topology), which is induced from the product
topology on $M^M$ where $M$ is equipped with the discrete topology.

In particular, if $M$ is a graded vector space, each $ \End_{\C}(M)_{\mu}
$ is a closed vector subspace of $ \End_{\C}(M)$, although
$\End_{gr}{M} $ is not closed in general. A graded vector subspace
$E=\oplus_{\mu}E_{\mu}$ of $ \End_{\C}(M)$ is called graded closed
in $ \End_{\C}(M)$ if each $ E_{\mu}$ is closed.

We define the topology on $A(g)$ to be the weakest topology such
that all induced maps $\rho_M$: $A(g)\to \End_{\C}(M)$ for $M\in
{\cal C}_g$ are continuous.

For each $ m \in M_{\mu}$, the left ideal
$\an_{A(g)}(m)=\{a\in A(g) | am=0\}$ of $ A(g)$ is a
graded left ideal of $ A(g)$. In fact, if
$a\in A(g)$ such that $ a = \oplus a_{\lambda}$ (a finite
sum with $a_{\lambda} \in A(g)_{\lambda} $ and distinct $ \lambda$'s), then
$ am=0  $ implies $ a_{\lambda}m=0 $ for all $ \lambda$, thus
$ a_{\lambda} \in \an_{A(g)}(m)$.

Let $\LL_g$ be the collection of  left ideals
of $A(g)$ defined by:
$$\LL_g=\{\an_{A(g)}(m)|m\in M, M\in {\cal C}_g\}.$$

 For each
$ J \in \LL_g $, we denote $ J_{\lambda}=J \cap (A(g))_{\lambda} $.
Then $ J\supseteq \oplus_{\lambda}J_{\lambda}$. Note that
$ \oplus_{\lambda}J_{\lambda}$ is  a graded left ideal of $U_g(V)$.
(In general, $ J $ is not graded.)

\begin{lem}
(1) For each $J\in \LL_g,$ $A(g)_n=J_n$ if $n\in \frac{1}{K}\Z$ is
sufficiently small.

(2) Each $A(g)_n/J_n$ is finite dimensional.

(3) For $J_1,J_2\in \LL_g$, there exists $J\in \LL_g$ such that
$J\subset J_1\cap J_2.$

(4) For any $a\in A(g)$ and $J\in \LL_g$ there exists $I\in \LL_g$ such that
$Ia\subset J.$

(5) $\cap_{J\in \LL_g}J=\{0\}$.

(6) Every $ J \in \LL_g $ contains a graded left ideal $ I \in \LL_g $.
\end{lem}

\pf For convenience we write $J(m)=\an_{A(g)}(m)$ for $m\in M$ and $M\in\CC_g.$

(1).  We assume, at first, that $J = J(m) $ for a
homogeneous element $ m \in M_{\lambda}$ and $M\in\CC_g.$
Consider the $A(g)$-module homomorphism:
$$\begin{array}{ccc}
A(g) & \to & A(g)m\\
a & \mapsto & am.
\end{array}$$
Then $A(g)m$ is a $ g$-twisted $ V $-submodule of $M$ by
Lemma~\ref{sec1.3}. Since $(A(g)m)_{\lambda+n}=A(g)_{n}m$
where $n\in\frac{1}{K}\Z$ and
 $(A(g)m)_{n+\l}=0$ for $n\in\frac{1}{K}\Z$ sufficiently small. Thus
we must have
$A(g)_n\subset J_n$ for $ n \in \frac{1}{K}\Z $ and sufficiently small.

In general, let $m=\sum_{\mu}m_{\mu}$ with $m_{\mu}\in M_{\mu}$ with
distinct $\mu$'s in $ \frac{1}{K}\Z $.
Since $J=\an_{A(g)}(m)\supseteq \cap_{\mu}J(m_{\mu})$. Note that
$ J_{\lambda}\supseteq \cap_{\mu}J(m_{\mu})_{\lambda}$ for any $ \lambda $.
Thus we have a surjective  linear map
$ A(g)_{\lambda}/\cap_{\mu}J(m_{\mu})_{\lambda}\rightarrow
A(g)_{\lambda}/J_{\lambda} $ and
$ (A(g))_{\lambda}/\cap_{\mu}J(m_{\mu})_{\lambda}\subseteq \oplus_{\mu}
 (A(g))_{\lambda}/J(m_{\mu})_{\lambda} $. Now the result follows from
 the proof for homogeneous $m $ above and the fact that there are
only finitely many nonzero $ m_{\mu}$'s. The proof of (2) is also
 included in this proof.

(3) Let  $m_i\in M^i$  such that $J_i=\an_{A(g)}(m_i)$ for $i=1,2.$
Consider the module $M=M^1\oplus M^2$
and $m=m_1+m_2.$  Then $a\in\an_{A(g)}(m)$ if and only if
$a\in J_1\cap J_2.$ That is, $J_1\cap J_2=\an_{A(g)}(m)\in \LL_g.$

(4) Let $J=\an_{A(g)}(m)$ with $m\in M$. For any $ a \in A(g)$, set
$I=\an_{A(g)}(am)\in \LL_g.$
It is obvious that $Ia\subset J.$

(5) If $ a \in \cap_{J\in\LL_g}J $, then $ \rho_M(a)= 0 $ for all
$ M $ in $ {\cal C}_{g} $. Therefore $ a= 0 $ in $ A(g) $ by the
definition of the algebra $A(g) $ (see \ref{sec:ag}).

(6) Let $ J=\an_{U_g(V)}m $ for some $ m \in M $ for a $ g $-twisted
$ V $-module $ M $. Let $ m=\sum_{i=1}^{s}{m_{i}} $ with $ m_{i}\in
M_{\lambda_{i}} $
as homogeneous components. We have $ J \supseteq \cap_{i=1}^{s}
\an_{A(g)}(m_{i})$.
Here $ \cap_{i=1}^{s}\an_{A(g)}(m_{i})=\an_{A(g)}((m_{1},\ldots, m_{s}))$,
with $ (m_{1},\ldots, m_{s}) \in \oplus_{i=1}^{s}A(g)m_{i} $,
is a graded left ideal and in
$ \LL_g $.
 \qed

\begin{th} The collection $\LL_g$ forms a basis at $0$ of a linear
(Hausdorff)  topology
on $A(g)$ such that the topology  defined earlier on $A(g)$ coincides with
this topology. In particular, the graded left ideals in $ \LL_g $ also
form a basis at $  0 $  of this linear topology.
\end{th}

\pf For each $ M $ in $ {\cal C}_{g} $, the pointwise convergence
topology on $ \End(M) $, which is a topological group which discrete
topology on $ M $,
 has a basis at 0 consists of sets
\[ W(m)=\{ f \in \End(M) \; | \; f(m)=0 \} \]
for any $ m \in  M $ and the finite intersection of these sets.
Since the topology on $ A(g) $ defined earlier is
 the weakest topology on $ A(g) $ such that $ \rho_M$:
$A(g)\to \End(M) $ is continues,
then $ \rho^{-1}_{M}(W(m))$ with $ m \in M $ and
$ M $ running through $ {\cal C}_{g} $ generates a basis of this
topology at 0 on $ A(g)$.
However, by definition one sees that
$  \rho^{-1}_{M}(W(m))=\an_{A(g)}(m) $. Therefore
$ \LL_g $ is a basis of this topology.

It is a direct consequence of Lemma \ref{sec1.4}(5) that
the topology is Hausdorff.
By Lemma  \ref{sec1.4}(6), the collection of graded left ideals
in $ \LL_g $ also form a basis of this topology at 0. \qed

\begin{cor} With the topology defined above the left and right multiplications
by an element of $A(g)$ are continuous linear maps.
\end{cor}

\pf Fix $a\in A(g).$ Clearly we have  $aJ\subset J$ for any $J\in{\cal L}_g.$
This shows that the left multiplication by $a$ on $A(g)$ is continuous.
By Lemma \ref{sec1.4} (4) there exists $I\in \LL_g$ so that $Ia\subset J,$
i.e., the right  multiplication by $a$ is continuous.
\qed

\subsect{Extending the category ${\cal C}_g$ by the
topology on $A(g).$}\label{sec1.5}
For each $ A(g)$-module $ M$, we define
$$\F(M)=\{m\in M|Jm=0 \ \text{\rom{for\ some\ }} J\in \LL_g\}.$$
First of all, by Lemma~\ref{sec1.4}(3), $ \F(M) $ is vector subspace of
$ M $. Further by  Lemma~\ref{sec1.4}(4), we see that for any
$ m \in \F(M)$, then $ a m \in  \F(M)$. Therefore $ \F(M) $ is an
$ A(g)$-submodule of $ M $. Following the definition, we can see
$\F(\F(M))=\F(M)$. In particular by the definition of $ \LL_g $,
we also have $ \F(M) = M $ for all $ M $ in $ {\cal C}_{g} $.

Note that $ A(g) $ is an associative algebra and is equipped with
the topology defined above. We say an $A(g)$-module $M$ is continuous if
the map $\pi$: $A(g)\times M \to M $ is continuous.
All modules are equipped with
discrete topology. However, an $A(g)$-module $M$ is continuous if and only if
$ \F(M) = M $. Indeed, all modules $ M$ such that $\F(M)=M $ is continuous
since for each $m \in M $, $(1+J)\times \{ m \} $ is an open set in
$ A(g)\times M $ and  $(1+J) m= m$. Here $ J \in \LL_g $ and $Jm=0 $.
Conversely,  if $ M $ is continuous, $ \pi^{-1}(0) $ is an open set in
$A(g)\times M$. For any $ m\in M $,
since $(0,m) \in \pi^{-1}(0) $, there must be a $ J \in \LL_g $ such that
$J\times \{ m \} \subseteq \pi^{-1}(0) $. This shows that $Jm=0$.

We now define $ \bar{{\cal C}}_{g} $ to be the
full subcategory of $ A(g) $-modules $ M $ such that  $ \F(M)=M $.
Therefore $ {\cal C}_{g}$ is a full subcategory of $ \bar{{\cal C}}_{g}$.
We use $ {\cal I}$:  $\bar{{\cal C}}_{g}\to A(g)$-\M\  to denote the
 natural embedding functor.
It is straightforward from the definition
to verify that $A(g)$-submodules and quotient of
modules in $ \bar{{\cal C}}_{g} $ are still in $ \bar{{\cal C}}_{g}$.
Thus for each $ A(g)$-module $ M $,
$ \F(M) $ is the unique maximal $ A(g) $-submodule of $ M $ contained
in $ \bar{{\cal C}}_{g} $. Furthermore,
if $ f$: $M \to N $ is a homomorphism of $ A(g) $-modules for any
two $ A(g) $-modules $ N $ and $ M $, then
$ \F(M)\subseteq \F(N) $. Therefore we see that
$ \F$: $A(g)$-$\M \to \bar{{\cal C}}_{g} $ is a functor.

\begin{lem} (1) $ ({\cal I}, \F ) $ is an adjoint pair of functors.

(2) $ \bar{{\cal C}}_{g} $ is an Abelian category and closed under the
direct limits of $A(g)$-modules.
\end{lem}
\pf (1) Let $ M \in \bar{{\cal C}}_{g} $ and $ N  $ any $ A(g) $-module.
Any $ f \in \Hom_{A(g)}({\cal I}M,N) $ if and only if $ f(M)\subseteq
\F(N) $. The verification of functorial property is routine.

(2) All other conditions of an Abelian category are straightforward.
We only verify that direct limits of $ A(g)$-modules in
$ \bar{{\cal C}}_{g} $ are still in $ \bar{{\cal C}}_{g}$.
Let $( M_{\alpha}, f_{\alpha,\beta})_{\alpha, \beta\in I } $
be  a direct system. Let $ M $ be the direct limit in $ A(g)$-Mod with
maps $  f_{\alpha}$: $M_{\alpha}\to M $. Note that as $ A(g) $-module,
$ M $ is generated by the submodules $ f_{\alpha}(M_{\alpha}) $ for all
$ \alpha \in I $. Since $ M_{\alpha} $ is in $ \bar{{\cal C}}_{g} $,
so is $  f_{\alpha}(M_{\alpha}) $. Thus $ f_{\alpha}(M_{\alpha}) \subseteq
\F(M) $ for all $\alpha \in I $. But $ \F(M) $ is a submodule of $ M $.
 Hence $ \F(M)=M $ and $ M $ is in $ \bar{{\cal C}}_{g} $.
\qed


\begin{th} Let $ M $ be any $ A(g)$-module and $ m \in\F(M)$.
Then $A(g)m $ is in $ {\cal C}_{g}$. In particular, every
module in  $ \bar{{\cal C}}_{g} $ is a union of
 submodules in $ {\cal C}_{g} $ and direct sum of $ L(0)$-eigenspaces.
\end{th}


\pf Since $m\in \F(M)$ there exists $J\in \LL_g$ such that $Jm=0.$
Let $W\in \CC_g$ and $w\in W$ with $J=\an_{A(g)}(w).$ Clearly,
$A(g)w$ is a $g$-twisted $V$-submodule of $W$.
Note that the map from $A(g)w$ to $A(g)m$ by sending
$aw$ to $am$  is well defined and
 an $A(g)$-module homomorphism.  Thus $A(g)m$ is
a $g$-twisted $V$-module, that is, $A(g)m\in\CC_g.$ The other assertions in
the theorem are clear now.
\qed

\begin{cor} (1) A module $ M $ in $ \bar{{\cal C}}_g $ is in $ {\cal C}_g $ if
and only if for each $ \lambda \in \C $,
$ M_{\lambda} $  is finite dimensional and
 $ M_{\lambda-n}= 0 $ for $ n \in \frac{1}{K}\Z $ with $ n >> 0 $.

(2) $ \ext_{{\cal C}_{g}}^{1}(M,N)=\ext_{\bar{\cal C}_{g}}^{1}(M,N)$ for
all $ M, N $ in ${\cal C}_{g}$.

(3) Every simple module in $ \bar{\cal C}_{g} $ is in $ {\cal C}_{g} $.
\end{cor}
\pf (1) Let $ M $ be any module in $\bar{{\cal C}}_g $. We can define
\[ Y_M(v, z)=\sum_{n \in \frac{1}{K}\Z}v_{n}z^{-n-1} \]
in $ \End(M)\{z\}$ for each $v \in  V $. Here $ v_{n} $ is the image
of
$ v\otimes t^n $ in $ \End(M)$.  Note that all conditions for
a $g$-twisted $V$-module except Eq.\ (\ref{bound1}) and Eq.\ (\ref{jacobi})
are satisfied. Now we show that for so defined $Y_M(v, z)$,
the Jacobi identity is also satisfied. We only need to verify Eq.\
(\ref{jacobi-term}). For each $ u \in V^{r} $ and $v \in V $, we denote by
\[ f(X)=\sum_{m\geq 0}(-1)^m
\comb{p}{m}(u_{p+s-m}v_{t+m}-(-1)^{p}v_{p+t-m}u_{s+m})X^m \]
\[g(X)=\sum_{m\geq 0}\comb{s}{m}(u_{p+m}v)_{s+t-m}X^m \]
to be two elements of $C$. We will use $f(X) $ and $ g(X) $ to denote their
images
in $A(g) $ as well.  Since, for any $g$-twisted $V$-module $W$, we have
$ f(X)w=g(X)w$ for all $ w \in W $ by the Jacobi identity (\ref{jacobi-term}),
we have $ f(X)=g(X)$ in $A(g)$. So we have $f(X)m=g(X)m $ for all $m \in M $
since $ M$ is an $ A(g)$-module. This only shows that $ f(X)=g(X)$ as
elements of $ A(g) $, but there no reason to  believe that the
actions of $ f(X) $ on $ M $ is the same as the ``limit'' as required
by  the Jacobi identity. Thus we have to show that for any $ m \in M $,
\[ f(X)m=\sum_{i= 0}^{n}(-1)^i
\comb{p}{i}(u_{p+s-i}v_{t+i}-(-1)^{p}v_{p+t-i}u_{s+i})m \]
and
\[ g(X)m=\sum_{i= 0}^{n}\comb{s}{i}(u_{p+i}v)_{s+t-i}m \]
for sufficiently large $n$.  Set
\[ f_n(X)=\sum_{i= 0}^{n}(-1)^i
\comb{p}{i}(u_{p+s-i}v_{t+i}-(-1)^{p}v_{p+t-i}u_{s+i})X^i \]
and
\[  g_n(X)=\sum_{i= 0}^{n}\comb{s}{i}(u_{p+i}v)_{s+t-i}X^i. \]
Here we have to use the fact that all modules in $ \bar{{\cal C}}_g $
are continuous. Since $M$ is in $\bar{{\cal C}}_g $, there exists
$ J \in \LL_g $ such that $ Jm=0 $. Then $J=\an_{A(g)}(w)$ for some $W $ in
${\cal C}_g $
and $ w \in W$. Since $ W $ is $ g $-twisted, there exists
$ n_0$ such that $ f(X)w=f_n(X)w $ for all $ n \geq n_0 $, i.e.,
$f(X)-f_n(X) \in J $. Thus we have $f(X)w=f_n(X)w $. Equality
$g(X)m=g_n(X)m $ for sufficiently large $n$ is proved in the same way.

Finally, in order for $ M$ in $ \bar{{\cal C}}_g $  to be in $ {\cal C}_g $,
one only needs the condition Eq.~(\ref{bound1}). But this is assumed in the
corollary.

(2) Let
$ 0 \to N \to E \to M \to 0 $ be  an extension in $ \bar{\cal C}_{g} $.
Since $ E $ is in $\bar{{\cal C}}_{g} $,  by (1),
we have  $ E \in {\cal C}_{g} $.  This shows that every extension
in $ \bar{{\cal C}}_{g} $ is actually an extension in
$ {\cal C}_{g} $. Since $ {\cal C}_{g}$ is a full subcategory of
$ \bar{\cal C}_{g} $, two extensions in $ {\cal C}_{g} $ are equivalent
$ {\cal C}_{g} $ if and only if  they are equivalent in
$ \bar{{\cal C}}_{g} $. (3) follows from the Theorem since $ A(g)m $ is a
submodule of $ M $ and in $ {\cal C}_{g}$,
\qed

\subsect{Construction of modules in $\bar{\cal C}_g.$}\label{sec1.6}
Let $M$ be a $\C $-vector space. Then space $\text{\rom{Hom}}_{\C}(A(g),M)$ has
an $A(g)$-module structure defined in the following way:
$$(af)(x)=f(xa)$$
for $f\in \Hom_{\C}(A(g),M)$ and $a\in A(g).$ One can easily see  that
$$(abf)(x)=f(xab)=(bf)(xa)=(a(bf))(x).$$
We define
$\X(M)=\F(\Hom_{\C}(A(g),M))$ then $\X(M)\in\bar{\cal C}_g.$

There is a natural $ \C $-linear map  ev: $\X(M)\to M $
defined by $ \text{\rom{ev}}(f)=f(1) $ for all $ f \in \X(M)$.
If $ N $ is another vector space $ \phi$: $N \to M $ is a linear map,
the natural induced map $
\tilde{\phi}$: $\Hom_{\C}(A(g),N)\to \Hom_{\C}(A(g),M)$
is an $ A(g)$-module homomorphism, which further induces an
$ A(g) $-module homomorphism $ \X(\phi)$: $\X(N)\to \X(M) $.
Thus one can verify that $ \X $ defines a functor
$\C $-Vect$\to \bar{{\cal C}}_{g},$ where $\C$-Vect is the category of
complex vector spaces.
If $ {\cal G}$: $\bar{{\cal C}}_{g}\to \C $-Vect is the forgetful functor,
then

\begin{lem} $\X$ is a left exact functor from the category of
vector spaces to $\bar{\cal C}_g$. Furthermore,
$ (\cal G, \X) $ is an adjoint pair.
\end{lem}

\pf As the right adjoint to the exact functor $\I,$
$\F$ is obvious left exact. It is well known that the
functor $\text{\rom{Hom}}_{\C}(A(g),\cdot)$ is left exact. So $\X$ is a
composition of two left exact functors and is left exact.

The adjoint pair follows from the following isomorphism of vector spaces
for any $ N $ in $ \bar{{\cal C}}_{g} $  and vector space $ M:$
\begin{eqnarray}
\Hom_{\C}(N,M)&=&\Hom_{\C}(A(g)\otimes_{A(g)} N,M)\nonumber\\
&\cong & \Hom_{A(g)}( N, \Hom_{\C}(A(g),M))= \Hom_{A(g)}( N, \X(M)).
\end{eqnarray}
Here the second isomorphism  comes from the adjointness of the
``tensor functor'' and ``$ \Hom $  functor'' while the last
equality follows from the fact that   $ N $ is  in $ \bar{{\cal C}}_{g}$.
Naturality of the isomorphism is also routine to check.
\qed

\begin{rem}
Since both $\F$ and $\X$ have  exact left adjoint functors, they send
injective modules to injective modules. In particular,
$ \X(M) $ is an injective module in $ \bar{{\cal C}}_{g}$ for $M\in\C$-Vect.
\end{rem}

  Let $M\in\bar{\cal C}_g.$ Define a linear map
$\phi$: $M\to \text{\rom{Hom}}_{\C}(A(g),M)$
by $\phi(m)=f_m$: $a\mapsto am$ for $m\in M$ and $a\in A(g).$
Then for $a,b\in A(g)$ we have
$$\phi(am)(b)=f_{am}(b)=b(am)=f_{m}(ba)=(af_m)(b)=(a\phi(m))(b).$$
This shows that $\phi$ is an $A(g)$-module homomorphism.\,Thus
$\phi(M)\!\subset\!\F(\text{\rom{Hom}}_{\C}(A(g),M))$
$=\X(M)$ since $ \phi(M) $ is in $ \bar{{\cal C}}_{g}$.
Next we show that $\phi$ is injective. If $\phi(m)=0$ then
$\phi(m)(a)=am=0$ for all $a\in A(g).$ In particular,
$\phi(m)(1)=m=0.$  Note that the functor
$ \X $ preserves the arbitrary direct sums. Then
$ \X(M)=\oplus_{b \in B }\X(\C b) $ with  $ B \subseteq M $ being a
$ \C $-basis. Thus we have proved

\begin{th} ${\cal C}_g$ has enough injective objects and every injective
module is a direct summand of  the direct sum of copies of $ \X(\C) $.
\end{th}

\subsect{}\label{sec1.9}
\begin{prop}
If $ {\cal C}_g $ is semisimple, then all modules in $ \bar{{\cal C}}_g $
are also semisimple.
\end{prop}
\pf Let $ M $ be a module in $ \bar{{\cal C}}_g $. For any $ m \in
M $, there exists a submodule $ N \subseteq M $ such that $ N $ is in
$ {\cal C}_g $ and $ m \in N $. Since $ N $ is a direct sum of simple
modules in $ {\cal C}_g $, we have shown that $ M $ is the sum of
all simple submodules in $ {\cal C}_g $. Thus $ M $ is a direct sum of simple
modules in $ {\cal C}_g $. \qed

\esub

\section{Induction for a subalgebra}
\label{sec2}

\bsub
\subsect{}\label{sec2.1}
Let $V'$ be a vertex operator subalgebra  of $V$ such that
$g(V')=V'.$  Let $g'=g|_{V'}$ and $o(g')=K'.$ Then $K'|K$ and
each $g$-twisted $V$-module $M$ when restricted to $V'$ is a
$g'$-twisted $V'$-module. Let $ {\cal C}'_{g'} $ be the category all
$ g' $-twisted $ V' $-module. Then  we have a restriction functor
$$\Res_{\bar{\cal C}'_{g'}}^{\bar {\cal C}_{g}}:\ {\cal C}_g\to{\cal
C}'_{g'}.$$

Consider the enveloping algebras $A(g)$ and $A(g').$ The
natural embedding
$\bar V'=\oplus_{r=0}^{K'-1}(V')^r\otimes t^{r/K'}\C[t,t^{-1}]\to
\bar V$ induces an algebra homomorphism
$\psi$: $T(\bar V')\to T(\bar V)$ where
$(V')^r=\{v\in V'|g'v=e^{r2\pi i/K'}v\}.$ For each $M\in{\cal C}_g$
the composition
of the algebra homomorphism $\rho_M$: $T(\bar V)\to \End_{\C}(M)$ with
$ \psi $ produces an algebra  homomorphism $\rho_M'$: $T(\bar V')\to
\End_{\C}(M).$ Since $M|_{V'}$ is a
module for $U_{g'}(V')$ the kernel $I'$ of $T(\bar V')\to U_{g'}(V')\to 0$
is contained in $\Ker \rho'_M$ and $\psi(\Ker \rho_M')\subset\Ker \rho_M.$
Thus $\psi(I')\subset I=\cap_{M\in {\cal C}_g}\Ker\rho_M$
and $ \psi $ induces an algebra
 homomorphism $ U_{g'}(V')\to U_g(V) $, which is still denoted by
$\psi $.
\begin{th} With the above assumption, the algebra homomorphism
 $\psi$: $U_{g'}(V')\to
U_g(V)$ induces  a homomorphism $ \bar\psi$: $A(g')\to A(g) $ of associative
algebras such that
for each $M\in{\cal C}_g,$  $\Res_{\bar {\cal C}'_{g'}}^{\bar {\cal C}_{g}}(M)$
as $A(g')$-module factors through
$A(g),$ that is, $A(g')\to A(g)\to \End_{\C}(M).$
\end{th}
\pf The homomorphism $\psi$: $U_{g'}(V')\to U_g(V) $ induces an
algebra homomorphism  $ \psi[X]$: $U_{g'}(V')[[X]]\to U_g(V)[[X]] $.
It follows  from the definition of $ C_{M}(g) $ in \ref{sec:ag}, that
$ \psi[X](C_{M|_{V'}})\subseteq C_{M} $. Since $ M|_{V'} $ is in
$ {\cal C}'_{g'}(V') $ for each  $ M $ in $ {\cal C}_{g}(V) $,
$ \psi[X] $ induces an   algebra homomorphism
$\bar{\psi}$: $A(g') \rightarrow A(g) $. The conditions on the
module structures in the theorem are clear from the construction of the
algebra homomorphism $\bar{ \psi} $. \qed

\subsect{}\label{sec2.2}
Let $\LL'_{g'}$ be the corresponding collection of  left ideals defining
the topology on $A(g')$ as in \ref{sec1.4}.
\begin{lem} For any $J\in{\cal L}_g$ there exists $J'\in\LL'_{J'}$ such that
$\bar\psi(J')\subset J.$
\end{lem}

\pf Let $m\in M$ with $M\in{\cal C}_g$ and $J=\an_{A(g)}(m).$ Set
$J'=\an_{A(g')}(m).$ Then $J'\in\LL'_{g'}.$ Now the result
follows.
\qed

Let $ {\cal C}'_{g'}$ be the
corresponding category for $ V'$ and $ g' $.
For any $ A(g') $-module $M' $, let $ \F'(M') $ be
the submodule of $ M' $ for $ A(g') $.  We will use $ \bar{{\cal C}}'_{g'}$
to denote the category of $ A(g')$-modules $ M' $ such that $ \F'(M')=M' $.
For each $ A(g) $-module $ M $, $ M|_{A(g')} $ is the
$ A(g) $-module via $ \bar{\phi} $. The lemma implies that
$ \F(M)\subseteq \F'( M|_{A(g')}) $.  In particular, $ M|_{A(g)} $
is in $ \bar{\cal C}'_{g'} $ if $ M $ is in $ \bar{\cal C}_{g} $.
We thus defines a functor $ \bar{{\cal C}}_{g}\to \bar{{\cal C}}'_{g'} $. We
will still use  $ \Res_{\bar {\cal C}'_{g'}}^{\bar {\cal C}_{g}} $ to denote
this functor
if $ V $ and $ V' $ are understood from the context.

\begin{prop} The restriction functor $\Res_{\bar {\cal C}'_{g'}}^{\bar {\cal
C}_{g}}$: $\bar{\cal C}_g\to \bar{\cal C}'_{g'}$
is exact.
\end{prop}

\subsect{}\label{sec2.3} We consider $A(g)$ as a left $A(g')$-module
and a right $A(g)$-module, i.e., $A(g)$ is a
$A(g')$-$A(g)$-bimodule. For each $M\in \bar{\cal C}'_{g'},$
$\Hom_{A(g')}(A(g),M)$ is a left
$A(g)$-module with the action $(af)(x)=f(xa)$ for
$a,x\in A(g)$ and $f\in  \text{\rom{Hom}}_{A(g')}(A(g),M).$ We define
$$\Ind_{\bar{\cal C}'_{g'}}^{\bar{\cal
C}_{g}}(M)=\F(\text{\rom{Hom}}_{A(g')}(A(g),M))
.$$
Then $\Ind_{\bar{\cal C}'_{g'}}^{\bar{\cal C}_{g}}(M)$ is a module in
$\bar{\cal C}_g.$

If $M$ and $N$ are in $\bar{\cal C}'_{g'}$ and $\phi$: $M \to N$ is a
homomorphism of $g'$-twisted $V'$-modules, then
$$\begin{array}{cccc}
\tilde\phi: & \text{\rom{Hom}}_{A(g')}(A(g),M) &\to
&\text{\rom{Hom}}_{A(g')}(A(g),N)\\
{} & f &\mapsto & \phi\circ f
\end{array}$$
is a homomorphism of $A(g)$-modules. Indeed, for  $f\in
\text{\rom{Hom}}_{A(g')}(A(g),M),$  and $a,x\in A(g)$, we have
$$(\tilde{\phi}(af))(x)=(\phi\circ(af))(x)=\phi(f(xa))=(\phi\circ f)(xa)
=(a\tilde{\phi}(f))(x),$$
that is, $\tilde{\phi}(af)=a\tilde{\phi}(f).$ Also, for $a'\in A(g')$
$$(\tilde\phi(f))(\bar{\psi}(a')x)=\phi(f(\bar{\psi}(a')x))=\phi(a'f(x))=a'\phi(f(x))=a'(\tilde{\phi}(f)(x)),$$
i.e., $\tilde{\phi}(f)\in \text{\rom{Hom}}_{A(g')}(A(g),N).$ Then
the map $f\to \phi\circ f$ induces a homomorphism
$$\Ind_{\bar{\cal C}'_{g'}}^{\bar{\cal C}_{g}}(\phi):\ \Ind_{\bar{\cal
C}'_{g'}}^{\bar{\cal C}_{g}}(M)\to \Ind_{\bar{\cal C}'_{g'}}^{\bar{\cal
C}_{g}}(N)$$
in $\bar{\cal C}_g.$ We thus define a functor $\Ind_{\bar{\cal
C}'_{g'}}^{\bar{\cal C}_{g}}$: $\bar{\cal C}'_{g'}\to \bar{\cal C}_g,$
which we call the induction  functor.

\subsect{}\label{sec2.4} For each $M\in\bar{\cal C}'_{g'}$, there is an
evaluation
map
$${ev}: \ \Ind_{\bar{\cal C}'_{g'}}^{\bar{\cal C}_{g}}(M)\to M$$
defined by $ev(f)=f(1).$

\begin{lem} $ev$ is a homomorphism of $A(g')$-modules when
$\Ind_{\bar{\cal C}'_{g'}}^{\bar{\cal C}_{g}}(M)$ is restricted to $A(g')$ via
$\bar{\psi}.$
\end{lem}

\pf For $a'\in A(g')$ and $f\in \Ind_{\bar{\cal C}'_{g'}}^{\bar{\cal
C}_{g}}(M)$, we have
$$ev(a'f)=(a'f)(1)=f(1\bar{\psi}(a'))=f(\bar{\psi}(a')1)=a'f(1)=a'ev(f),$$
as desired.
\qed

\begin{lem}
For any $M\in\bar \CC_g$ we have
$ev: \ \Hom_{U_g(V)}(U_g(V),M)\simeq M.$ Moreover, if $M\in \CC_g$
then $\Ind_{\bar {\cal C}_{g}}^{\bar {\cal C}_{g}}(M)\simeq M.$
\end{lem}

\pf
We need to prove that
$ev$ is one to one and onto in this case.
If $f(1)=0$ then $f(a)=af(1)=0$ for any
$a\in U_g(V)$ thus $f=0$ and $ev$ is one to one. Let $m\in M$  and define
$f_m\in \Hom_{U_g(V)}(U_g(V),M)$ by $f_m(a)=am.$   Then $f(1)=m$ and $ev$ is
onto.
\qed

\begin{th}[Frobenius Reciprocity] For any $E\in\bar{\cal C}_g$ and
$M\in\bar{\cal C}'_{g'}$ the natural map
$$\Phi:\  \text{\rom{Hom}}_{A(g)}(E,\Ind_{\bar{\cal C}'_{g'}}^{\bar{\cal
C}_{g}}(M))\to
\text{\rom{Hom}}_{A(g')}(\Res_{V'}^V(E),M)$$
defined by $\phi\mapsto ev\circ\phi$ is an isomorphism of
vector spaces.
\end{th}

\pf First $ev\circ \phi=\Phi(\phi)$ is a composition of
$A(g')$-homomorphisms and thus is a homomorphism
in $\bar{\cal C}'_{g'}.$ We only  need to construct
the inverse $\Phi^{-1}.$ For each $\phi'\in\
\Hom_{A(g')}(\Res_{\bar {\cal C}_{g'}'}^{\bar {\cal C}_{g}}(E),M)$
define
$$\phi:\  E\to \text{\rom{Hom}}_{\C}(A(g),M)$$
such that
$$(\phi(e))(x)=\phi'(xe)$$
for $e\in E$ and $x\in A(g).$ In fact $\phi$ is the composition:
$$\begin{array}{ccccc}
A(g) & \to & E &\to & M\\
x &\mapsto & xe& \mapsto & \phi'(xe).
\end{array}$$
Then $\phi(e)$  is a homomorphism of $A(g')$-modules. Since
for $x'\in A(g')$
$$\phi(e)(\bar{\psi}(a')x)=\phi'(\bar{\psi}(a')xe)=a'\phi'(xe)=a'\phi(e)(x).$$
Thus we have a map $\phi$: $E\!\to\!\text{\rom{Hom}}_{A(g')}(A(g),M).$
Moreover $\phi$ is an $A(g)$-homomorphism:
$$ \phi(ae)(x)=\phi'(xae)=\phi'(e)(xa)=(a\phi(e))(x).$$
Since $E\in \bar{\cal C}_g$ so is its image $\phi(E)$ in
$\text{\rom{Hom}}_{A(g')}(A(g),M).$ Now by the definition of the functor
$\F$ we have $\phi(E)\subset \Ind_{\bar{\cal C}'_{g'}}^{\bar{\cal C}_{g}}(M).$
We define $\Phi^{-1}(\phi')=\phi.$ Clearly $\Phi^{-1}$ is a linear
map.

To finish the proof, we need to show that $\Phi^{-1}$ is an inverse of
$\Phi.$ Let \\$f\in\text{\rom{Hom}}_{A(g)}(E,\Ind_{\bar{\cal
C}'_{g'}}^{\bar{\cal C}_{g}}(M))$ and $\phi'=
ev\circ f.$
Then
$$\phi(e)(x)=\phi'(xe)=(ev\circ f)(xe)=f(xe)(1)=(xf(e))(1)=f(e)(x)$$
for $e\in E$ and $x\in A(g)$. Thus $\phi(e)=f(e),$ $\phi=f$
and $\Phi^{-1}\circ \Phi=id.$

Conversely, if $\phi'\in
\text{\rom{Hom}}_{A(g')}(\Res_{\bar{\cal C}'_{g'}}^{\bar{\cal C}_{g}}
(E),M)$ and $\phi=\Phi^{-1}(\phi'),$ then
$$\Phi(\phi)(e)=(ev\circ \phi)(e)=ev(\phi(e))=\phi(e)(1)=
\phi(e)(1)=\phi'(1e)=\phi'(e),$$
i.e., $\Phi\circ \Phi^{-1}=id.$ This completes the proof.
\qed

\begin{cor} (1) $(\Res_{\bar{\cal C}'_{g'}}^{\bar{\cal C}_{g}}
,\Ind_{\bar{\cal C}'_{g'}}^{\bar{\cal C}_{g}})$ is an adjoint pair
of functors.

(2) $\Ind_{\bar{\cal C}'_{g'}}^{\bar{\cal C}_{g}}$ is left exact and sends
injective modules to
injective modules. \qed
\end{cor}
\begin{rem}
For a module $ M $ in $ \bar{\cal C}_{g} $, define
$ \text{\rom{Soc}}_{\bar{\cal C}_{g}}(M) $, called the socle of $ M $,
 to be  the largest
semisimple submodule of     $ M $ and $ \text{\rom{Hd}}_{{\cal C}_{g}}(M)
$, called the head of $ M $,  to be the largest semisimple quotient of
$ M $. If $ M $ a semisimple module
(need not be finite length) and $ S $ is a simple module, we denote
by $ [M, S] $ the cardinality of the copies of $ S $ appearing in the
direct sum decomposition into simple modules.
If $ M $ is a simple module in $\bar{{\cal C}}_{g'} $ and $ N $
simple module in  $\bar{{\cal C}}_{g} $, then the Frobenius reciprocity
implies that
\[ [ \text{\rom{Hd}}_{{\cal C}'_{g'}}(\Res_{\bar {\cal C}'_{g'}}^{\bar {\cal
C}_{g}}(N)):M]=
[ \text{\rom{Soc}}_{\bar{\cal C}_{g}}(\Ind_{\bar {\cal C}'_{g'}}^{\bar {\cal
C}_{g}}(M)):N].\]
However, it is not clear that in the category $ {\cal C}_{g} $,
every nonzero module has a nonzero socle or quotient.
\end{rem}

\subsect{Transitivity.}\label{sec2.5} Let $V''\subset V'\subset V$
be subalgebras of $V$ which are $g$-stable. Set $g''=g'|_{V''}.$
We shall use the obvious notation
$\bar{\cal C}_{g''}''$ for the corresponding category of $ g''
$-twisted $ V'' $-modules.    Let $M\in \bar{\cal C}''_{g''}$ and define
$$\xi_M: \ \Ind_{\bar {\cal C}''_{g''}}^{\bar {\cal C}_{g}}(M)\to
\Ind_{\bar{\cal C}'_{g'}}^{\bar{\cal C}_{g}}\Ind_{\bar{\cal
C}''_{g''}}^{\bar{\cal C}'_{g'}}(M)
$$
by $\xi_M(f)=\bar f$: $A(g)\to
\Ind_{\bar {\cal C}''_{g''}}^{\bar {\cal C}'_{g'}}M.$ The definition of $\bar
f$
is as follows: for $x\in A(g),$ $\bar f(x)$ is a map:
$$\begin{array}{cccc}
\bar f(x):\ & A(g')&\to &M\\
{} & y &\mapsto & f(\bar\psi(y)x).
\end{array}$$
Recall that $\bar\psi$ is the algebra homomorphism from $A(g')$
to $A(g).$

\begin{th} $\xi_M$ is an isomorphism of $A(g)$-modules, and further
induces an isomorphism of functors $\xi$: $\Ind_{\bar {\cal
C}''_{g''}}^{\bar {\cal C}_{g}}\to
\Ind_{\bar{\cal C}'_{g'}}^{\bar{\cal C}_{g}}
\Ind_{\bar{\cal
C}''_{g''}}^{\bar {\cal C}'_{g'}}.$
\end{th}

\pf First we show that $\xi_M(f)\in \Ind_{\bar{\cal C}'_{g'}}^{\bar{\cal
C}_{g}}\Ind_{\bar {\cal C}''_{g''}}^{\bar {\cal C}'_{g'}}(M).$
Let $\psi'$ and $\psi''$ be the induced algebra homomorphisms
$\psi'$: $A(g'')\to A(g')$ and
$\psi''$: $A(g'')\to A(g),$ respectively. Then
$\psi''=\bar\psi\circ \psi'$. For $z\in A(g'')$ and  $f\in \Ind_{\bar {\cal
C}''_{g''}}^{\bar {\cal C}'_{g'}}(M)$,
we have
$$\bar f(x)(\psi'(z)y)=f(\bar\psi(\psi'(z)y)x)=f(\psi''(z)\bar\psi(y)x)
=zf(\bar\psi(y)x)=z\bar f(x)(y),$$
i.e., $\bar f(x)$ is an $A(g'')$-module homomorphism.

To
see $\bar f(x)\in\Ind_{\bar {\cal C}''_{g''}}^{\bar {\cal C}'_{g'}}(M)$, we
need to show that there
exists $J'\in\LL'_{g'}$ such that $J'\bar f(x)=0.$ Let $J\in \LL_g$ such that
$Jf=0$ or equivalently $f(J)=0.$ From Lemma \ref{sec1.4} (4) there
exists $I\in{\cal L}_g$ so that $Ix\subset J.$ By Lemma \ref{sec2.2} there
exists $J'\in\LL'_{g'}$ such that $\bar\psi(J')\subset I.$ Thus
$$(J'\bar f(x))(y)=\bar f(x)(yJ')=f(\bar\psi(yJ')x)\subset f(Ix)\subset
f(J)=\{0\},$$
that is, $\bar f(x)\in\Ind_{\bar {\cal C}''_{g''}}^{\bar {\cal C}'_{g'}}(M).$

For $a'\in A(g')$, we have
$$\bar f(\bar\psi(a')x)(y)=f(\bar\psi(y)\bar\psi(a')x)=f(\bar\psi(ya')x)=\bar
f(x)(ya')=
(a'\bar f(x))(y)$$
where $y\in A(g')$ is arbitrary. This shows that
$\bar f\in \text{\rom{Hom}}_{A(g')}(A(g),\Ind_{\bar {\cal C}''_{g''}}^{\bar
{\cal C}'_{g'}}(M)).$

In order to prove  $\bar f\in \Ind_{\bar {\cal C}'_{g'}}^{\bar {\cal
C}_{g}}\Ind_{\bar {\cal C}''_{g''}}^{\bar {\cal C}'_{g'}}(M)$ it is
enough to show that $J\bar f=0$ where $J\in \LL_g$ such that $Jf=0.$
Note that $(J\bar f)(x)=\bar f(xJ)=0$ for all $x\in A(g)$
iff $\bar f(J)=0.$ But $\bar f(J)(y)=f(\bar\psi(y)J)\subset f(J)=0$ for
all $y\in A(g')$. Thus $\bar f(J)=0.$

Next we show that $\xi_M$ is a homomorphism of $A(g)$-modules.
Let $a\in A(g)$ and $f,$ $x,y$ be as before. Then
$$\xi_M(af)(x)(y)=(af)(\psi(y)x)=f(xa)(y)=(a\xi_M(f))(x)(y),$$
or $\xi_M(af)=a\xi_M(f).$

It remains to show that $\xi_M$ is an isomorphism of $A(g)$-modules.
We achieve this by constructing an inverse $\eta$ of $\xi.$
For $\phi\in  \Ind_{\bar {\cal C}'_{g'}}^{\bar {\cal C}_{g}}\Ind_{\bar
{\cal C}''_{g''}}^{\bar {\cal C}'_{g'}}(M)$,  define
$\eta(\phi)=f$: $A(g)\to M$ by $f(x)=ev\circ \phi(x)=\phi(x)(1)$ (note that
$\phi(x)\in \Ind_{\bar {\cal C}''_{g''}}^{\bar {\cal C}'_{g'}}(M)$).
It is easy to verify that
$f\in \text{\rom{Hom}}_{A(g'')}(A(g),M).$ In fact if $a''\in A(g'')$
then
\begin{eqnarray*}
f(\psi''(a'')x)&=&\phi(\psi''(a'')x)(1)=(\psi'(a'')\phi(x))(1)\\
&=&\phi(x)(\psi'(a''))
=a''(\phi(x)(1))=a''f(x).
\end{eqnarray*}
Since there exists $ J \in \LL_g $ such that $ J\phi = 0 $, then
 $$(Jf)(x)=f(xJ)=\phi(xJ)(1)=0 .$$
Then $ f \in \Ind_{\bar{\cal C}''_{g''}}^{\bar{\cal C}_{g}}M $.
Thus $\eta_M(\phi)=f$ yields  a linear map from
$\Ind_{\bar {\cal C}'_{g'}}^{\bar {\cal C}_{g}}\Ind_{\bar
{\cal C}''_{g''}}^{\bar {\cal C}'_{g'}}(M)$ to
$\Ind_{\bar {\cal C}''_{g''}}^{\bar {\cal C}_{g}}(M)$. Moreover, $\eta$ is
a homomorphism of $A(g)$-modules:
$$\eta_M(a\phi)(x)=(a\phi)(x)(1)=\phi(xa)(1)=f(xa)=(af)(x).$$

In order to finish the proof we need to check that
$\xi_M\circ \eta_M=id$ and $\eta_M\circ \xi_M=id.$ Let
$f\in \Ind_{\bar {\cal C}''_{g''}}^{\bar {\cal C}_{g}}(M)$ and
$\phi\in \Ind_{\bar{{\cal C}}'_{g'}}^{\bar{{\cal
C}}_{g}}\Ind_{\bar{{\cal C}}''_{g''}}^{\bar{{\cal C}}'_{g'}}(M).$
Then
$$((\eta_M\circ\xi_M)f)(x)=(\eta_M\bar f)(x)=\bar f(x)(1)=f(x),$$
$$((\xi_M\circ\eta_M)\phi)(x)(y)\!=\!(\overline{\eta_M \phi})(x)(y)\!
=\!(\eta_M
\phi)(\bar\psi(y)x)\!=\!\phi(\bar\psi(y)x)(1)\!=
\!(y\phi(x))(1)\!=\!\phi(x)(y)$$
for all $x\in A(g)$ and $y\in A(g').$ The proof is complete.
\qed

\subsect{}\label{sec2.6} Since $\Ind_{\bar{\cal C}'_{g'}}^{\bar{\cal
C}_{g}}$ is left exact and the
category $\bar{\cal C}'_{g'}$ has enough injective objects we can
consider the right
derived functors $R^i\Ind_{\bar{\cal C}'_{g'}}^{\bar{\cal C}_{g}}.$
For each $g'$-twisted $V'$-module
$M$ in $\bar{\cal C}'_{g'},$ $R^i\Ind_{\bar{\cal C}'_{g'}}^{\bar{\cal
C}_{g}}(M)$ is a $g$-twisted $V$-module in
$\bar{\cal C}'_{g'}.$ Furthermore, since  $\Ind_{\bar{\cal
C}'_{g'}}^{\bar{\cal C}_{g}}$
sends  injective modules to injective modules, we have the Grothendieck
spectral sequences with $E_{2}$-terms
$$R^i\Ind_{\bar{\cal C}'_{g'}}^{\bar{\cal C}_{g}}\circ
R^j\Ind_{\bar{\cal C}''_{g''}}^{\bar{\cal C}'_{g'}}(M)\Rightarrow
R^{i+j}\Ind_{\bar{{\cal C}}''_{g''}}^{\bar{{\cal C}}_{g}}(M)$$
for each $g''$-twisted $V''$-module $M$ in $\bar{\cal C}''_{g''}.$

\subsect{}\label{sec2.7} We have already defined
 the functor $\X$ from the category of
vector spaces to $\bar{\cal C}_g$ and proved it to be  exact earlier.
Now we can see that $\X$ can be defined with the induction functor.
 Note that our induction even makes sense
if we take $A$ being any subalgebra of $A(g)$ such that
the corresponding category $\bar{\cal C}_A$ is defined in the following way:
an $A$-module $M$ is in $\bar{\cal C}_A$ if and only if
$\{m\in M|(J\cap A)m=0\ \text{\rom{for\ some }}\ J\in \LL_g\}.$
In particular we may take $A=\C$.  Then $ \bar {\cal C}_A =
\C$-Vect because for any $J\in{\cal L}_g$ the intersection
$J\cap\C=\{0\}$. Thus
$$\Ind_{\bar{\cal C}_A}^{\bar{\cal C}_g}(M)=\X(M).$$

\esub

\section{The $g$-rational vertex operator algebras}
\label{sec4}
\bsub
\subsect{}\label{sec4.1}
In this section, we consider the induced modules for
$g$-rational vertex operator algebras. We call $ V $  $g$-rational,
if $ {\cal C}_{g} $ is semisimple with finitely irreducible modules.

However, we still fix an arbitrary  vertex operator algebra $ V $ and an
automorphism $ g $
of finite order $ K $ of $ V $. On each $ g $-twisted
$ V $-module, the operator $ L(0) $ acts
semisimply and the  graded $ A(g) $-module structure is
given by the eigenspaces of the operator $ L (0) $.
Similar to a finite dimensional semisimple Lie algebra, we say
$ \lambda \leq \mu $ for two weights $ \lambda , \mu \in \C $  if
$ 0\leq \mu -\lambda \in \frac{1}{K}\Z$.

Recall that the algebra $ A(g) $ is graded with gradation given by the
 adjoint action of $ L(0) $. The weight space $ A(g)_0 $ is an associate
subalgebra of $ A(g)$, with the operator $ L(0) $ being in the center
of $ A(g)_0 $.
On every finite dimensional simple  $ A(g)_0 $-module,
$ L(0) $ acts as a scalar.

The subspace
$ A^{\leq 0}=\oplus_{0 \leq n \in \frac{1}{K}\Z }A(g)_{-n} $ is a graded
subalgebra of $ A(g) $, and
$A^{-}=\oplus_{0 < n \in \frac{1}{K}\Z }A(g)_{-n} $ is a graded ideal of
$ A^{\leq 0} $.  Let $ M $ be a
module in $ {\cal C}_{g} $, an weight $ \lambda $ of  $ M $ is called
a minimal weight if there is no weight of $ M $ that is smaller than $\l$
in the sense defined above. Note that each weight space  $ M_{\mu}$  of $ M $
is an
$ A(g)_{0} $-module. However, if $ \lambda $ is a minimal weight of
$ M $, then $ M_{\lambda} $ is an $ A^{\leq 0}$-module with $ A^- $
acting  as zero.

\begin{lem} Let $ M $ be an irreducible module in $ {\cal C}_{g}$, then
there is a unique minimal weight $ \lambda $ such that
$ M_{\lambda} $ is a finite dimensional  simple $ A(g)_{0}$-module.
\end{lem}

\pf By the definition, there exists $ \lambda \in \C $ such that
$ \lambda $ is minimal and $ M_{\lambda}$ is not zero. If $ \mu $ is
a different minimal weight with   $M_{\mu}\neq 0 $,
 then $ A(g)M_{\mu}=\oplus_{ n \in
\frac{1}{K}\N}A(g)_{n}M_{\mu}\neq 0 $ is a submodule of $ M $, which does not
contain the subspace $ M_{\lambda} $. This contradicts the
irreducibility of $ M $. Thus $ \lambda $ is the only minimal
weight. If $0\neq N \subseteq M_{\lambda} $ is an $ A(g)_{0} $-submodule
of $ M_{\lambda}$, then $ A(g)N=\oplus_{ n \in
\frac{1}{K}\N}A(g)_{n}N $ is a submodule of $ M $
by the Jacobi identity (\ref{jacobi-term}), with $ p=0 $. However
we have  $  (A(g)N)_{\lambda} =N $. $N\neq 0 $ implies that
$ A(g)N \neq 0 $. The
irreducibility of $ M $ implies that $N=M_{\lambda}$.
 The  finite dimensionality
of $ M_{\lambda} $ follows from the definition of
modules in $ {\cal C}_{g} $. \qed

A module $ M $ in $ {\cal C}_g  $ is  called a lowest weight module
if there is a vector $ m \in M_{\lambda} $ for for a minimal weight
$ \lambda $ such that $ M = A(g)m $. Thus all  irreducible modules
are lowest weight modules by the lemma.  For  a
lowest weight module $ M $  of lowest weight $ \lambda $,
all weights of $ M $ are of the
form $ \lambda +\nu $ with $ \nu \in \frac{1}{K}\N $.

\subsect{} \label{sec4.2}
\begin{prop} Let $ M^1 $ and $ M^2 $ be two lowest weight
modules $ {\cal C}_{g} $ with lowest weights $ \lambda_1 $ and
$ \lambda_2 $  respectively.
If there exists a non-split extension
$$ 0 \rightarrow  M^1 \rightarrow E \stackrel{\pi}{\rightarrow}
  M^2 \rightarrow 0 $$
 in $ {\cal C}_g $, then $ \lambda_{1}-\lambda_{2} \in \frac{1}{K}\Z $.
\end{prop}

\pf Let $ m_2 \in M^2_{\lambda_2} $ such that $ A(g)m_2 =M^{2}$.
Since $ E $ is in $ {\cal C}_{g} $,
there exists $0 \neq m \in E_{\lambda_{2}} $
such that  $ \pi(m)=m_{2} $.
Then we  consider the  submodule $ A(g)m $ of $E $. All weights of $ A(g)m $
are in $ \lambda_{2} +\frac{1}{K}\Z $. If  $  A(g)m \cap M^1 = 0 $
then $ \pi$: $A(g)m \rightarrow M^2 $ is an isomorphism
since $ \pi(A(g)m )=M^2 $. This contradicts the nonsplitness of the
exact sequence. Thus we must have  $  A(g)m \cap M^1 \neq 0 $ which is
a $ g $-twisted $ V $-submodule of $ M^1$.
 We pick  $0\neq  x \in  A(g)m\cap M^1 $.
 We can assume that $ x $ is a weight vector of weight
$ \nu $. Thus $ \nu \in \lambda_1 +\frac{1}{K}\Z $. But we also have
 $ \nu \in \lambda_2 +\frac{1}{K}\Z $. This shows that
$ \lambda_1 -\lambda_2 \in \frac{1}{K}\Z $. \qed

\subsect{}\label{4.3}
Let $ N $ be a finite dimensional $ A(g)_{0}$-module such that
$ L(0) $ acts on it by a scalar $ \lambda $. Then we can extend
$ N $ to an $ A^{\leq 0 }$-module by letting $ A^-$ act as zero.
We consider the  $A(g)$-module
$ M(N)=A(g)\otimes_{A^{\leq 0}}N $. However, $ M(N) $ needs not be
in $ \bar{{\cal C}_g} $. The followings are some standard properties
of the module $ M (N)$:

(1) $ M(N) $ is a graded $ A(g) $-module with the gradation given by the
eigenspaces of the operator $ L(0)$ and the projection
$ \text{pr}_{\lambda}$: $M(N)_{\lambda}\rightarrow N $ is an isomorphism of the
vector
spaces (actually as $ A^{\leq 0}$-modules).

(2) All weights of $ M(N) $ are in $ \lambda+\frac{1}{K}\N $.

(3)  Suppose that $ N $ is a simple  $ A(g)_0 $-module. Then
there is a nonzero quotient of  $ M(N) $ in the category
$ {\cal C}_g $ if and only if  there is a simple module in the category
$ {\cal C}_g $ with lowest weight space isomorphic to $ N $ as
 $ A(g)_0 $-modules. Furthermore, if  $ M(N)$ has any
nonzero quotient  in $ {\cal C}_g $, then it has a unique maximal
submodule and thus a unique simple quotient in $ {\cal C}_g  $.
Here any $ A(g)$-submodule containing a weight vector of the lowest weight
$ \lambda $ contains the whole module $ M(N) $ since $ N $ is simple
while the sum of all submodules not containing
 the weight $\lambda $
is still such a submodule.

(4) Sharpiro's lemma holds: for any module $ M $ in $ \bar{\cal C}_g $
and any finite dimensional
$ A(g)_{0}$-module $ N $, we
have the canonical isomorphism
\[ \Hom_{A(g)}(M(N), M)=\Hom_{A(g)_0}(N, M_{\lambda}^{A^-}) .\]
Here $ M_{\lambda}^{A^-} $ is the subspace of $ M_{\lambda}$ consisting
all vectors killed by all elements of $ A^- $.

\rem It is not known in general the $ A(g) $-module $ M(N)$ has a
unique maximal quotient in $ \bar{{\cal C}}_g $,
 which we would like to denote by $ D(N) $.
However we  expect that this is the case. If the module $ D(N)$ exists and not
zero,
then in the above listed properties we can simply replace $ M(N) $ by $ D(N)$
so that all the above listed properties hold when $ M(N) $ is replaced
by $ D(N) $ except the isomorphism of the projection in (1).

\subsect{}\label{sec4.5}
\begin{prop}
Let $ M^1 $ and $ M^2 $ be two irreducible modules in $ {\cal C}_g $
with lowest weights $ \lambda_1 $ and  $ \lambda_2 $ respectively.
Then $ M^1 $ and $ M^2 $ are isomorphic in $ {\cal C}_g $ if and only if
$ M_{\lambda_{1}} $ and $ M_{\lambda_2}$ are isomorphic as $ A(g)_0 $-modules.
\end{prop}
\pf One direction of the proposition is clear. We only show that
$ M^1 $ and $ M^2 $ are isomorphic if $ M_{\lambda_1} $ and
$ M_{\lambda_2}$ are isomorphic as $ A(g)_{0} $-modules.
We first observe that that $ \lambda_1=\lambda_2  $ (we denote them by
$ \lambda $)  since they are the scalars
by which the operator $ L(0) $ acts on the two modules respectively.
    Let $ N^1=M_{\lambda}^1 $ and $ N^2=M^2_{\lambda}$. Then $ N^1 $
and  $ N^2 $ are irreducible $ A(g)_{0}$. We consider
the $ A(g)$-module $ M(N^1) $. By (4) of the previous subsection,
we have
\[ \Hom_{A(g)}(M(N^1), M^2)=\Hom_{A(g)_0}(N^1,N^2)\neq 0  .\]
Thus $ M(N^{1}) $ has a unique simple quotient, which has to be  isomorphic to
$ M^2 $ since $ M^{2} $ is simple. We can replace $ M^{2} $ by $ M^{1} $
in the above argument to show that $ M^1 $ is also a simple quotient of
$ M(N^{1}) $. Thus $ M^1 \cong  M^2 $ by the uniqueness of the simple
quotient of $ M(N^1) $. \qed

\rem This proposition is related to some results obtained in [DLM]
(also see [Z]). In [DLM]
an associative algebra $A_g(V)$ associated with the vertex operator algebra
$V$ and the automorphism $g$ is constructed. It is proved there that
there is a bijection between the set of inequivalent
weak irreducible $g$-twisted modules (the eigenspace of $L(0)$ in a weak
module can be infinite-dimensional) and the set of inequivalent irreducible
$A_g(V)$-modules. One can easily verify that $A_g(V)$ is a quotient
of $A(g)_0.$


\subsect{}\label{sec4.7}
Recall that a block $ {\cal B} $ in $ {\cal C}_g $ is a full subcategory
of $ {\cal C}_g $ generated by an equivalent class of irreducible
modules with respect to the equivalent relation generated by
the following: $ M^1 $ and $ M^2 $ are equivalent if there is
a non-split extension in $ {\cal C}_g $ between them.
By Lemma \ref{sec4.2}, each block uniquely determines an element
in $ \C /(\frac{1}{K}\Z) $ by taking the images of the lowest weights of
the irreducible modules in the block. We say that a block is bounded above if
 $ \lambda+\frac{1}{K}\Z $ is the class in $ \C/(\frac{1}{K}\Z) $ corresponding
to
the block $ {\cal B} $ and the set of lowest weights
 of  all irreducible modules in  $ {\cal B} $ is a subset of
$ \lambda+\frac{1}{K}\Z $ with an  upper bound.

\begin{prop}
Suppose that all blocks in $ {\cal C}_g $ are bounded above,
 then every $A(g)$-finitely generated  module in $ {\cal C}_g $ has a
composition series.
\end{prop}
\pf Let $ M $ be a finitely generated module in $ {\cal C}_g $. Without
lose of generality, we can assume that $ M $ is generated by finitely many
weight vectors, say, $ m_1, \ldots, m_n $ with weights $
\lambda_1, \ldots , \lambda_n $. Consider the finite filtration of
submodules generated by these vectors:
\[0=M^0\subseteq  M^1=\langle m_1 \rangle \subseteq \langle m_1, m_2 \rangle
 \subseteq \cdots \subseteq M^n=\langle m_1, \ldots, m_t \rangle=M .\]
We see that   each subquotient $ M^i/M^{i-1} $ is generated by a single
weight vector. Thus we can assume that $ M $ is generated by
a single vector.  Thus all weights of $ M $ are of the form
$ \lambda+\frac{1}{K}\Z $. Furthermore since all the weight spaces are finite
dimensional and only finitely many weight spaces of  weights $
\lambda - \frac{j}{K}$ with $ j \in \N $. By a similar argument,
we can assume that the generator is a lowest weight vector.
Thus $ M $ is a lowest weight module. Now we need to use the
condition that every block is bounded above. Let $ n $ be the positive integer
such that there is no irreducible module of lowest weights
of the form $ \lambda+\frac{j}{K} $ with $ j > n $. We can choose
$ n=n_{\lambda} $ as the smallest for the each such fixed $ \lambda $
and use induction on $ n_{\lambda} $. If $ n_{\lambda}=0 $, we note
that
$ M_{\lambda} $ is finite dimensional. Take $ N\subseteq M_{\lambda} $
as an irreducible $ A(g)_0 $-submodule. Consider the $ A(g)$-module
$ M(N) $. We have a non-zero homomorphism
$ \phi$: $M(N) \rightarrow M $ resulted from the embedding
$ N \rightarrow M_{\lambda}$. Then $M(N) $ has a nonzero quotient in
$ {\cal C}_g $ and the unique maximal submodule $M(N)_m $
has all weights of the form $\lambda+\frac{j}{K} $ with $ j >0 $.
If the image of $ \phi $ is not irreducible, it has to have a submodule
with all weights of the form $ \lambda+\frac{j}{K} $ with $ j > 0 $.
It has to have a lowest weight vector which generates a submodule with a
simple quotient in $ {\cal C}_g $ with lowest weight larger than
 $ \lambda $.  This contradicts the assumption that $ n_{\lambda}=0 $.
Now we can use a induction on the dimension of $ M_{\lambda} $ and
reduce the proof to  the case such that $ M_{\lambda} $ is irreducible as $
A(g)_0 $-module. A similar argument shows that $ M $ has to be simple in this
case. For $ n_{\lambda}>0 $, We can use a similar argument to find
submodules $F\subseteq E $   of $M $ such that  $ E/F $ is irreducible
with lowest weight $ \lambda $ and both $ F $ and $ M/E $ have
either larger lowest weight or same lowest weight but strictly
smaller dimension of the lowest weight space. By induction we
see that both $ F $ and $ M/E $ have composition series. Thus $ M $
has a composition series.
\qed

\begin{cor}
With the conditions in as in the Theorem, all modules in $ {\cal C}_g$
 with weights in a finite union $ \cup_{i=1}^n (\lambda_{i}+\frac{\Z}{K})
$
have composition series.
\end{cor}
Actually the proof of the Theorem works for such assumption.

\subsect{}\label{sec4.8}
\begin{prop}
If all blocks in $ {\cal C}_g $ are bounded above, then any non-zero
module in $ {\cal C}_g $ has non-zero simple submodules
and non-zero simple quotient.
\end{prop}
\pf Consider for each $ \lambda \in \C $ the subspace $
M(\lambda)=\sum_{i\in \Z}M_{\lambda+\frac{i}{K}}$ which, in fact, is a
submodule of $ M $. By
Lemma \ref{sec4.2}, we must have $ M =
\oplus_{\lambda \in \C/\frac{\Z}{K}} M(\lambda)$.
By Corollary \ref{sec4.7}, each $ M(\lambda) $ has a composition
 series if it is not zero. Thus each non-zero $ M(\lambda) $ has a
 non-zero simple submodule and quotient.
\qed

\subsect{}\label{sec4.9}

\begin{lem}
Let $ V' $ be a $ g $-invariant vertex subalgebra of $ V $ and
$ g'=g|_{V'} $. If $ {\cal C}'_{g'} $ is bounded above, so is $ {\cal C}_g $.
\end{lem}

\pf Let $ M $ be a simple module in $ {\cal C}_g $. Then $ M $ is
a module in $ {\cal C}'_{g'} $. The lowest weight space $ M_{\lambda} $
generates a submodule in $ {\cal C}'_{g'}$, which has a simple subquotient of
lowest weight
$\lambda $. That the $ {\cal C}'_{g'} $ is bounded above implies that
$ {\cal C}_g  $ is necessarily bounded above.
\qed

\subsect{}\label{sec4.11}
A natural  question about the induction functor is
when $ \Ind_{\bar {\cal C}'_{g'}}^{\bar {\cal C}_{g}}N $ is not zero
for an irreducible module $N $ in
$ {\cal C}'_{g'} $. Since we are interested in modules in $ {\cal C}_{g} $
only. So another natural question is whether  the
module $ \Ind_{\bar {\cal C}'_{g'}}^{\bar {\cal C}_{g}} N $ is  in $
{\cal C}_g $ whenever $N $ is in
$ {\cal C}'_{g'} $. We discuss  these questions in the
following special case.
 We start with
the following lemma.
\begin{lem}
Let $ L $ and $ M $ be  modules in $ {\cal C}_g $  such that $ L $
is finitely generated.  Then $ \Hom_{{\cal C}_g}(L, M) $ is finite
dimensional.
\end{lem}

\pf Since $ L $ is finitely generated, we can
further assume that  the finite set of generators are
all weight vectors. Let $ N $ be the subspace generated
by the generators, which is a finite direct sum of finite dimensional
weight spaces.
    Let $ \phi$: $L \rightarrow M $ be any homomorphism in $ {\cal C}_g $.
Then  $ \phi $ is uniquely determined by
the restriction map $\phi_{0}$: $N\rightarrow M_{0}$. Here $ M_{0} $ is the sum
of the
weight subspaces of $ M $ corresponding to all the weights that appear in
$ N $.
Now both $ N $ and $ M_{0} $ are finite dimensional.
\qed.

\rem A similar argument shows that if $ M $ has a composition
series, then $ \Hom_{{\cal C}_g}(L,M)$ is finite dimensional for all
$ L $ in $ {\cal C}_g $.
This can be reduced to the case that $ M $ is irreducible. Then one can
use the fact that any module in $ {\cal C}_g $  can not
have a quotient which is a direct
sum of an irreducible module infinitely many times due to the finite
dimensionality  of the weight spaces of modules in $ {\cal C}_g $.

\begin{th} Let $V$ be a $g$-rational vertex operator algebra.
Then $ \ind_{\bar {\cal C}'_{g'}}^{\bar {\cal C}_{g}}L $ is in $ {\cal
C}_g $ for  every irreducible
module $ L $ in $ {\cal C}'_{g'} $. More generally, $  \ind_{\bar {\cal
C}'_{g'}}^{\bar {\cal C}_{g}}N $ is
in $ {\cal C}_g $ for any module $ N $ in $ {\cal C}'_{g'}  $ which has a
composition series.
\end{th}
\pf First of all, since $ {\cal C}_g $ has only finitely many irreducible
modules, then any module in $ {\cal C}_g $ has a composition series.
Thus a module in $ \bar{{\cal C}}_{g} $ is non-zero if and only if
it has a non-zero simple submodule.  Take any simple module $ S $ in
$ {\cal C}_g $. By the Frobenius reciprocity, we have
$ \Hom_{\bar{\cal C}_g}(S,
\ind_{\bar {\cal C}'_{g'}}^{\bar {\cal C}_{g}}L)=\Hom_{{\cal C}'_{g'}}(S,L) $.
Since $ L $ is simple in $ {\cal C}'_{g'} $, $ \Hom_{{\cal C}'_{g'}}(S,L)$
is finite dimensional following the above remark. Thus
$ S $ appears in $  \ind_{\bar {\cal C}'_{g'}}^{\bar {\cal C}_{g}}L $
only finitely many times  as direct summand
since the category $ {\cal C}_g $ is semisimple (thus $ \bar{{\cal
C}}_{g} $ is also semisimple). Since
$ {\cal C}_g $ has only finitely many irreducible modules,
$  \ind_{\bar {\cal C}'_{g'}}^{\bar {\cal C}_{g}}L $ is a direct sum
of  finitely many irreducible modules and thus in $ {\cal C} $. \qed
\esub

\section{Inductions for simple vertex operator algebras}\label{sec5}
\bsub
\subsect{}\label{sec5.1}
In this section we discuss the case when $ V $ is a simple vertex
operator algebra.  Fix an automorphism $ g $ of finite order
$ K $ of $ V $. Then $ V =\oplus_{r=0}^{K-1}V^{r} $ is a direct sum
decomposition of  $ g $-eigenspaces. It is proved in
\cite{dm:gal} that $ V'=V^0 $ is a simple
vertex operator algebra and $ V^{r} $ is an irreducible $ V^0 $-module.
Note that for any $ r $, $ u \in V^r $ homogeneous,
 and any $ g $-twisted module $ M $ in $ {\cal C}_g $, the  degree of
 the operator $ u_{n}$  (as a linear operator on the graded vector space
$ M $) equals $ \text{wt}(u)-n-1 $ with $ n \in \frac{r}{K}+\Z $.
In particular, for each $ u \in V^0 $, $ u_{n} $ has degree in $ \Z $.
Thus  for any $ g $-twisted $ V $-module $ M $ and
$ \lambda \in \C $ the vector space
$M^0(\lambda)=\oplus_{n \in \Z} M_{\lambda+n} $ is a $ V^0 $-submodule of
$ M $. Note that  $ M^0(\lambda)=M^0(\mu) $ if and only if
$ \lambda -\mu \in \Z $.
On the other hand, $ M(\lambda)=\oplus_{n \in \frac{1}{K}\Z}M_{\lambda+n}
$ is an $ g $-twisted $ V $-submodule of $ M $ and $ M(\lambda)=M(\mu) $
if and only if $ \lambda -\mu \in \frac{1}{K}\Z $. Therefore,
we have  $ M(\lambda)=\oplus_{r=0}^{K-1}M^0(\lambda+\frac{r}{K}) $, which
gives a  decomposition as $V^0 $-modules.

For each fixed $ \lambda $, we can define a $\langle g \rangle $-action
on $ M(\lambda) $
such that $ g $ acts on $ M^{0}(\lambda-\frac{r}{K}) $ by
the scalar $\xi^{r} $. Here we fix $ \xi= e^{\frac{2\pi i}{K}}$.

\begin{lem} With the above action, we have
\[ g\circ Y_{M(\lambda)}(u,z)\circ g^{-1} =Y_{M(\lambda)}(gu,z) \]
for all $ u \in V $.
\end{lem}
\pf Let $ u \in V^r $ and $ m \in M^{0}(\lambda+\frac{j}{K}) $
 be homogeneous elements.
For each $ n \in \frac{r}{K}+\Z $, $ u_{n}(m)$ has weight
$ \wt\,v+\wt\,m-n-1  \in \wt\,m-\frac{r}{K} +\Z $. Thus
$ u_{n}(m) \in M^0(\lambda-\frac{j}{K}-\frac{r}{K})$ and
\[ g(u_{n}(g^{-1} m))=g(\xi^{-j} u_n(m))=\xi^{-j} \xi^{j+r} u_{n}(m)
=\xi^r u_{n}(m)=(\xi^{r}u)_{n}m=(gu)_{n}m .\]
\qed

\rem Through the decomposition
$ M=\oplus_{\lambda \in \C/(\frac{\Z}{K})}M(\lambda) $,
 $ g $ acts on $ M $. However, this action is not functorial
and depends on the  choice of the $ \lambda $ in $ \lambda
+\frac{\Z}{K} $.

\subsect{}\label{sec5.2}
Let $ h $ be an automorphism of $ V $, commuting with $ g $. Then $ h $
preserves the $ g $-eigenspace decomposition of $ V $.
Let $ M $ be any $ g $-twisted module with
the vertex operators denoted by $ Y_{M}(v,z) $.
Let $ ^hM =M $ as vector spaces. For each $ v \in V $, we define
$ Y_{^hM}(v,z)=Y_{M}(hv,z) $.
It is straightforward to verify that $ Y_{^hM}(v, z) $ makes $ ^hM $
into a $ g $-twisted $ V $-module.  If $ \phi$: $M\to N $ is a
homomorphism of $ g $-twisted $ V $-modules, then $ ^h\!\phi=\phi$:
$^h\! M \to ^h \! N $ is also a homomorphism of $ g $-twisted $ V
$-modules.  Thus the assignment
$ M \mapsto \; ^h\!M $ defines a functor
$ {\cal C}_g \rightarrow {\cal C}_{g}$ and
$ ^{(h_{1}h_{2})}M=^{h_{1}}\!(^{h_{2}}M) $.

\begin{lem}
Let $ M $ be a $ g $-twisted $ V $-module, then for any choice of
the action $ g$: $M \rightarrow M $ defined in the remark above gives an
isomorphism $ M \rightarrow\; ^h\! M $ of $g$-twisted modules.
\end{lem}
\pf For each $ \lambda $, the action of $ g $ on $ M(\lambda) $ defines
an isomorphism of $ g $-twisted $ V $-modules by  Lemma \ref{sec5.1}. Then the
isomorphism extends to  $ M $ through the direct sum decomposition.
\qed

\subsect{}\label{sec5.3}
Recall that for each homogeneous $ u \in V^r $, the
weight of $ u_{n} $ is in $ \Z -\frac{r}{K} $ as operator on any
$  g $-twisted $ V $-modules.  We set $ A^{r}=\oplus_{n \in \Z}
A(g)_{n-\frac{r}{K}} $. Then we have
\[A(g)=\oplus_{r=0}^{K-1}A^{r} \]
which  gives  another graded algebra structure on  $ A(g) $.

Since for each automorphism $ h $ of $ V $ commuting with $ g $, the
module $ ^h M $ is also a $ g $-twisted module, one verifies directly from
the definition that $ h (u_{n})=(h u)_{n} $ defines an automorphism of
the   associated algebra $ A(g) $. In particular  $ g $ defines an
automorphism of $ A(g) $ such that the gradation
$ A(g)=\oplus_{r=0}^{K-1}A^{r} $ is given by the $ g $-eigenspaces
with $ g $ acts on $ A^r $ by $ \xi^{r} $.

Now let $ M $ be any module in $ {\cal C}_g $. The decomposition
$ M(\lambda)=\oplus_{r=0}^{K-1}M^0(\lambda-\frac{r}{K} ) $ is
a graded $ A(g) $-module with respect to the above gradation of $A(g)$.
This graded
$A(g)$-module structure is independent of  the choice of $ \lambda $
up to the index shifting.
 Moreover, for any  fixed $ g $ action on $ M $ compatible with the
$g$-action $ V$ in the sense of Lemma \ref{sec5.1}, the $ g $-eigenspace
decomposition of $ M $ gives  the  graded $ A(g) $-module structure on
$ M $.

We will concentrate on the case when $ M $ is simple.
If we assume the lowest weight spaces is contained in the $ g $-invariant
spaces of $ M $, then there is a unique way to index the homogeneous
spaces such that the lowest weight space is in $ M^0 $ and  we will
denote them by $ M=M^0\oplus M^1\oplus \cdots \oplus M^{K-1}$. This
notation is compatible   to the case when $ V $ is a simple $ V $-module (here
$ g = 1 $).

Our next goal is to show that if $ M $ simple, then $ M^0, M^1, M^{K-1} $ are
non-isomorphic
simple (untwisted) $ V^0 $-modules. In the following $ V $ is still an
arbitrary vertex operator algebra. The result is a strong contrast
with Lie algebra and indicates the associative algebra feature of
vertex operator algebras.

\subsect{}\label{sec5.4}
\begin{prop}
Let $ M $ be a  module in $ \bar{{\cal C}}_{g} $ and $ m \in M $ be
homogeneous.
Then the subspace $ \an_{V}(m)=\{ v \in V \; |\; Y_{M}(v, z)m= 0 \} $
is a $ V $-submodule of $ V $ (thus an ideal of $ V $).
\end{prop}
\pf Since $ m $ generates a submodule of  $ M $ in $ {\cal C}_g $.
We can assume that $ M $ is in $ {\cal C}_g $. Thus all weight spaces
are finite dimensional.
For any $ v \in V $, we can write $ v=v^0+\cdots +v^{K-1} $ such that
$ v^{r} \in V^r $. By considering the weights, we have
$ Y_{M}(v, z)m=0 $ if and only if $ Y_{M}(v^{r},z)m = 0 $ for all
$ r $. Thus $ \an_{V}(m)=\oplus_{r=0}^{K-1}(V^{r}\cap \an_{V}(m)) $.

Let $ M'=\oplus_{\lambda}M^*_{\lambda} $ be the
graded dual of $ M $. We need to discuss the associativity of the
vertex operator product expansion (cf. [DM2], [DL]).
 Let
$\C[z_1,z_2]_{S}$ be the localization of the polynomial ring $ \C[z_1,z_2] $
with respect to the set of all nonzero homogeneous polynomials of degree 1 and
$\iota_{i_1i_2}:  \C[z_1,z_2]_S \to
\C[[z_1,z_1^{-1},z_2,z_2^{-1}]]$  the injective map such that an element
$(az_1+bz_2)^{-1}$ is expanded in nonnegative integral powers of $z_{i_2}$
where $(i_1,i_2)$ is an  ordering of the set $\{1,2\}.$
Let $ v \in V^{i}\cap \an_{V}(m)  $
and $ u \in V^j $. For each $ m' \in M' $
by the rationality for  $ g $-twisted
modules, there exists $ f(z_{1}, z_2) \in \C[z_1,z_2]_{S}$
such that
\[ m'(Y_{M}(u,z_1)Y_{M}(v,z_2)m)z_{1}^{\frac{j}{K}}z_{2}^{\frac{i}{K}}
=\iota_{12}f(z_1,z_2)\]
and by the associatively, we have
\[ m'(Y_{M}(Y(u,z_0)v,z_2)m)(z_0+z_2)^{\frac{j}{K}}z_{2}^{\frac{i}{K}}
=\iota_{20}f(z_0+z_2,z_2)\]
for the same $ f $. Since $ v \in \an_{V}(m)$, we have  $ \iota_{12}f(z_1,z_2)=
0 $, which implies  $ f(z_{1},z_2)=0 $. Thus
$ f(z_{0}+z_{2},z_2)=0 $ and $ \iota_{20}f(z_0+z_2,z_2)=0 $.
Therefore
$$ m'(Y_{M}(Y(u,z_0)v,z_2)m) =0 .$$
Since $ m'\in M' $ is arbitrary, we have $ Y_{M}(Y(u,z_{0})v,z_{2})m=0 $.
Thus $ u_{n}(v)\in \an_{V}(m) $ and $ \an_{V}(m) $ is a
$ V $-submodule of $ V $. \qed

\begin{cor} Suppose  $ V $ is a simple vertex operator algebra.
Let $ M $ be a  module in $ {\cal C}_g $  with any fixed $ g $-action on
$ M $ such  the resulted eigenspace decomposition
$ M=\oplus_{r=0}^{K-1}M^{r} $ gives a graded $ A(g)$-module as
in \ref{sec5.3}. Then $ M\neq 0 $ implies $ M^{r} \neq  0 $.
\end{cor}
\pf We may assume that $ M $ is irreducible and the gradation is the
canonical one. Fix a nonzero vector $ m $ in the lowest weight space. Then
we have $ M=A(g)m =\oplus_{r=0}^{K-1}A^{r}m $ and
$ A^{r}m \subseteq M^{r}  $. Thus we have the equality $  A^{r}m= M^{r}$.
Note that $ u_{n} \in A^{r} $ for all $ n\in \frac{1}{K}\Z $ whenever
$ u \in V^r $. Suppose $ M^{r}=0 $ for some $ r $. Then we have
$ V^{r}\subseteq \an_{V}(m) $.
Thus the submodule $ \an_{V}(m) \neq 0 $ by Theorem 2 of \cite{dm:gal}.
However, $ {\bf 1} $ is not in $  \an_{V}(m) $. This contradicts the
 simplicity of $ V $. \qed

\subsect{}\label{sec5.5} Recall from \ref{sec1.3} that $ \bar{V}=
\oplus_{i=0}^{k-1}(V^i\otimes t^{i/K}\C[t,t^{-1}]) $ and  there
 is a natural
homomorphism $ \bar{V} $ to $ A(g) $. Thus $ \bar{V} $ ``acts'' on every
module in $ {\cal C}_g $. Note that $ \bar{V}$ is only a subspace in
$A(g)$. However the following explains that $ U_g(V) $ is not much
larger than $ \bar{V}$. The  proof of the lemma
uses a similar argument as in \cite{dm:gal} for the ordinary module
cases by using the rationality and associativity properties for
$ g $-twisted modules. For simplicity we denote $ \bar{V}^r=V^r \otimes
t^{\frac{r}{K}}\C[t,t^{-1}]$. Then $ \bar{V}=
\oplus_{r=0}^{K-1}\bar{V}^{r}$.

\begin{lem}
Let $V$ be any vertex operator algebra and $g$ is an automorphism of
finite order $K$ of $V$. If  $M$ is a $g$-twisted $V$-module generated by
a set $S$ over $V$, i.e., $ M=U_g(V)S$, then $M$ is the linear span of the
following subset
$\{ u_{n}s \; | \; u \in V, n \in \frac{\Z}{K}, s \in S \} $,
or simply $M=\bar{V}S $. \qed
\end{lem}

\begin{th}
Let $ V $ be a simple vortex operator algebra and $ M $ a simple
$ g $-twisted module. Then, in the  decomposition
$M=M^{0}+M^1+\cdots M^{K-1} $, $ M^{0}, \ldots , M^{K-1} $ are nonzero
and non-isomorphic simple $ V^{0} $-modules.
\end{th}
\pf By the corollary above, we know that $ M^{i}$'s are nonzero.
Next we show that they are irreducible.  Let $ 0\neq m \in M^{i} $.
Since $M$ is simple, thus $S=\{m\} $ is a generating set of $M$ on
$V$. By the lemma above, we have $M=\bar{V}=
\oplus_{r=0}^{K-1}\bar{V}^{r}S $.
However, $\bar{V}^{r}S  \subseteq M^{i-r}$. This implies that
we have the equality $M^{i-r}=\bar{V}^{r}S $. In particular,
we have $\bar{V}^{0}S=M^i$ and $ m$ generates the entire module $M^i $
over $V^{0}$.
This shows the simplicity of $M^i$.

To show that $M^0, \ldots ,M^{K-1} $ are non-isomorphic $V^0 $-modules,
one can follow a similar argument as in the proof of Theorem 4.1
of \cite{dm:gal}. \qed

\subsect{}\label{sec5.5a}
\begin{th}
Let $ V $ be a simple vertex operator algebra and $ g $ is an
automorphism of order $ K $. Suppose $ M $ and $ N $ are two
irreducible modules  in $ {\cal C}_{g}$ such that a $ V^0 $-component
$ M^r $ of $ M $ is isomorphic to a $ V^0 $-component $ N^s $ of $ N $.
Then $ M \cong N $ in ${\cal C}_g $.
\end{th}
\pf Let $ M=M^0+\cdots +M^{K-1} $ and $ N=N^0+\cdots +N^{K-1}$
be $ V^0 $-decompositions of $ M $ and $ N $ respectively. Suppose
$ \phi$: $M^r \rightarrow N^s $ is an isomorphism of $ V^0 $-modules.
Then $ \psi$: $M^r \rightarrow M^r\oplus N^s $ defined by
$ \psi(m)=(m,\phi(m)) $ is a $V^0 $-module homomorphism with image
isomorphic to $ M^r $  and a proper submodule of $ M^r \oplus N^s $.
We now fix $ 0 \neq m\in M^r $. Consider the
$ g $-twisted $ V $-submodule $ W $ generated by $ (m,\phi(m)) $ in
$ M\oplus N $. Then we have $ W=\bar{V}(m,\phi(m))=\sum_{i=0}^{K-1}
 \bar{V}^{i}(m,\phi(m)) $. However, $  \bar{V}^{i}(m,\phi(m))
\subseteq M^{r-i}\oplus N^{s-i} \subseteq M\oplus N $ for all
$ i $.
In particular for $ i=0 $, we have $ \bar{V}^{0}(m,\phi(m))=\psi(M^r)
 $ is a proper submodule of $ M^r\oplus N^s $. Therefore, $ W $ is a proper
submodule of $ M\oplus N $ and $ W $ must be a simple
submodule. Since  both  projections $ \mbox{pr}_1$:
$W \rightarrow M $ and $  \mbox{pr}_2$: $W \rightarrow N $ are nonzero,
they must be isomorphisms, which gives an isomorphism between $ M $
and $ N $ in $ {\cal C}_g $. \qed

\subsect{}\label{sec5.6}
Now we consider the induction from $V^{0} $ to $V$ for a simple vertex
operator algebra $ V $ with an automorphism  $ g $ of finite
order. Let $ {\cal C}^0_1 $ be the category of  untwisted $ V^0
$-modules.
\begin{prop}
Let $ M$ be any simple module in ${\cal C}_g$. Then
$ \ind_{\bar {\cal C}_{1}^0}^{\bar {\cal C}_{g}}M^r \neq 0 $. More precisely,
 $  \ind_{\bar {\cal C}_{1}^0}^{\bar {\cal C}_{g}}M^r$ has a unique
simple submodule isomorphic to $ M $.
\end{prop}
\pf Note that the projection $ M \rightarrow M^{r} $ is nonzero
homomorphism of $ V^0 $-modules. By the Frobenius reciprocity
we have a non-zero homomorphism $ M \rightarrow
\ind_{\bar {\cal C}_{1}^0}^{\bar {\cal C}_{g}}M^r$.
Since $ M $ is irreducible in $ {\cal C}_{g} $, $ M $ is isomorphic
to a submodule of $ \ind_{\bar {\cal C}_{1}^0}^{\bar {\cal C}_{g}}M^r
$. By Theorem \ref{sec5.5},
we have $\Hom_{{\cal C}_{1}^0}(N,M^r)=\C $. This shows that
$ M $ appears in the socle of $  \ind_{\bar {\cal C}_{1}^0}^{\bar
{\cal C}_{g}}M^r $ exactly once as a direct summand.
 On the other hand for any simple module $ N $ in $ {\cal C}_g $,
$\Hom_{{\cal C}_{1}^0}(N,M^r)\neq 0 $ if and only if $ N\cong M $
by Theorem \ref{sec5.5a}. \qed

\subsect{}\label{sec5.7}

\begin{th} Let $ V $ be a simple vertex operator algebra. Suppose that
${\cal C}_g $ is semisimple.

(1) Then for each simple $ V^0 $-module $ N $,
\[ \ind_{\bar{\cal C}_1^{0}}^{\bar{\cal C}_g} N \cong
\left \{ \begin{array}{rl} M &  \text{ if  }
 N \cong
M^{r} \text{ for a simple module $M$ in ${\cal C}_{g}$ and some $ r $},\\
 0 & \mbox{ otherwise;} \end{array}\right. \]

(2) If we further assume that $ {\cal C}_g $ has only finitely many
simple modules, then for any module $ N $ in ${\cal C}_1^0 $,
$ \ind_{\bar{\cal C}_1^{0}}^{\bar{\cal C}_{g}}N $ is in $ {\cal C}_g $.
\end{th}

\pf  By applying the Frobenius reciprocity and  using the
Theorem \ref{sec5.5}, we see that the identity of (1)
holds if the left hand side is replaced by its socle. Since $ {\cal C}_g$
is semisimple, then $ \ind_{\bar{\cal C}_1^{0}}^{\bar{\cal C}_{g}}N $ is also
necessarily semisimple by Proposition \ref{sec1.9}. Thus (1) is proved.

 For the same reason as above,
$  \ind_{\bar{\cal C}_1^{0}}^{\bar{\cal C}_{g}}N $ is semisimple. Since
for each simple module in $ {\cal C}_{g}$,  it has only finitely
many irreducible direct summands as  a $V^0 $-module.
On the other hand, for each  simple module $ L $ in $ {\cal C}_g $,
$ \Hom_{{\cal C}_1^0}(L, N ) $ is finite dimensional by
 Lemma~\ref{sec4.11}.
Therefore $\Hom_{{\cal C}_1^0}(M, N ) $ is finite dimensional as well for
any simple  $ g $-twisted $ V $-module $ M $
by Theorem \ref{sec5.5}.
Thus by the Frobenius reciprocity, each simple $ g $-twisted $ V $-module
appears only finitely many times as a direct summand of
$\ind_{\bar{\cal C}_1^{0}}^{\bar{\cal C}_{g}}N $.
Since $ {\cal C}_g $ has
only finitely many simple modules. Then
$  \ind_{\bar{\cal C}_1^{0}}^{\bar{\cal C}_{g}}N $ is
a direct sum of finitely simple modules and thus in $ {\cal C}_g $.
\qed

\rem  If $V$ is simple and rational (i.e., $g$-rational with $ g=1$),
the above theorem implies
that $ V=\ind_{\bar {\cal C}_{1}^0}^{\bar {\cal C}_{1}}V^0 $ if we
consider the ordinary modules.
The induction functor is actually defined
{}from ${\cal C}_1^0 $ to $ {\cal C}_g $.
  \esub

{\footnotesize

\noindent Address: {\sc Department of Mathematics,
 University of California,
 Santa Cruz, CA 95064}

\hspace{0.8cm} dong@dong.ucsc.edu
\vspace{0.4cm}

\hspace{0.8 cm}  {\sc Department of Mathematics, Kansas State University,
Manhattan, KS 66506}

\hspace{0.8cm} zlin@math.ksu.edu
}


\begin{thebibliography}{abcdef}
\bibitem[APW]{apw} H.H.\ Andersen, P.\ Polo, K.\ Wen, Representations
of quantum algebras, {\em Invent.\ Math.} {\bf 104} (1991), 1--59.
\bibitem[B]{B} R. E. Borcherds, Vertex algebras, Kac-Moody algebras,
and the Monster, {\em Proc. Natl. Acad. Sci. USA} {\bf 83} (1986), 3068-3071.
\bibitem[DPR]{1} R. Dijkgraaf, V. Pasquier and P. Roche,
Quasi-quantum groups related to orbifold models, Ed. by M. Carfora, M.
Martellini, A Marguolis, World Scientific, 1992, 75-89.
\bibitem[DVVV]{DVVV} R. Dijkgraaf, C. Vafa, E. Verlinde and H.
Verlinde, The operator algebra of orbifold models, {\em Comm. Math.
Phys.} {\bf 123} (1989), 485-526.
\bibitem[D1]{d1} C. Dong, Vertex algebras associated with even lattices,
{\em J. Algebra} {\bf 160} (1993), 245-265.
\bibitem[D2]{d} C. Dong, Twisted modules for vertex algebras associated
with even lattices, {\em J. Algebra} {\bf 165} (1994), 91-112.
\bibitem[D3]{d3} C. Dong,  Representations of the moonshine module
vertex operator algebra, {\em Contemporary Math.} {\bf 175} (1994).
\bibitem[DL]{dl:gen} C. Dong, J. Lepowsky, Generalized Vertex
Algebras and Relative Vertex Operators, {\em Progress in Math.} Vol. 112,
Birkh\"{a}user, Boston 1993.
\bibitem[DLM]{1} C. Dong, H. Li and G. Mason,  Twisted representations of
vertex operator algebras, preprint.
\bibitem[DM1]{dm:gal} C. Dong and G. Mason, On quantum Galois theory,
preprint.
\bibitem[DM2]{dm2:gal} C. Dong and G. Mason, On the operator content of
orbifold models, preprint.
\bibitem[DMZ]{dmz:gal} C. Dong, G. Mason and Z. Zhu,
Discrete series of the
Virasoro algebra and the moonshine module, {\em Proc. Symp. Pure. Math.,
American Math. Soc.} {\bf 56} II (1994), 295-316.
\bibitem[FFR]{ffr} A. J. Feingold, I. B. Frenkel and J. F. X. Ries, Spinor
construction of vertex operator algebras, triality and $E^{(1)}_8,$ {\em
Contemporary Math.} {\bf 121,} 1991.
\bibitem[FHL]{fhl} I. B. Frenkel, Y.-Z. Huang and J. Lepowsky, On
axiomatic approaches to vertex operator algebras and modules,
{\it Memoirs American Math. Soc.} {\bf 104}, 1993.
\bibitem[FLM]{flm} I. B. Frenkel, J. Lepowsky and A. Meurman,
Vertex Operator Algebras and the Monster, {\em Pure and Applied
Math.,} Vol. 134, Academic Press, 1988.
\bibitem[FZ]{1} I. Frenkel and Y. Zhu, Vertex operator algebras associated to
representations of affine and
Virasoro algebras, {\it Duke Math. J.} {\bf 66} (1992), 123-168.
\bibitem[HL]{HL} Y.-Z. Huang and J. Lepowsky, Toward a theory of
tensor products for representations of a vertex operator algebra, in: {\em
Proc. 20th Intl. Conference on Differential Geometric Methods
in Theoretical Physics, New York, 1991,} ed. S. Catto and A. Rocha,
World  Scientific, Singapore, 1992, Vol. 1, 344-354.
\bibitem[J]{ja} J.C.\ Jantzen, Representations of Algebraic
Groups, Academic Press, Orlando, 1987.
\bibitem[L]{li} H. Li, An approach to tensor product theory for
representations of a vertex operator algebra, Ph.D. thesis, Rutgers University,
1994.
\bibitem[Lin1]{lin1} Z. Lin,
  Induced representations of Hopf algebras:
applications to quantum groups at roots of 1,
{\em J. Algebra}, {\bf 154} (1993), 152-187.
 \bibitem [Lin2]{lin2} Z. Lin,
  A Mackey decomposition theorem and cohomology for the quantum groups
at roots of 1, {\em J. Algebra}, {\bf 154} (1993), 152--187.
\bibitem [V]{vogan} D.\ Vogan, Representations of Real Reductive
Lie Groups, Birkhauser, Boston-Basel-Stuttgart, 1981.
\bibitem[W]{1}
W. Wang, Rationality of Virasoro vertex operator algebras, {\it Duke
Math. J. IMRN}, {\bf Vol. 71}, No. 1 (1993), 197-211.
\bibitem[Z]{Z} Y. Zhu, Vertex operator algebras, elliptic functions and
modular forms, Ph.D. dissertation, Yale University, 1990.
\end{thebibliography}
\end{document}